# Sustainable Collaborative Strategy in Pharmaceutical Refrigerated Logistics Routing Problem

Tingting Chen[a], Feng Chu[b], Jiantong Zhang[a],*, Jiaqing Sun[a]

[a]School of Economics and Management, Tongji University, Shanghai 200092, China

[b]BISC, Univ Évry, University of Paris-Saclay, Évry, France

* Corresponding author.

E-mail address: zhangjiantong@tongji.edu.cn.

# Sustainable Collaborative Strategy in Pharmaceutical Refrigerated Logistics Routing Problem

The rapid growth of pharmaceutical refrigerated logistics poses sustainability challenges, including elevated costs, energy consumption, and resource inefficiency. Collaborating multiple depots can enhance logistics efficiency when standalone distribution centers have limited transport resources, i.e., refrigerated vehicles. However, the sustainable benefits and performance across different strategies remain unexplored. This study fills this research gap by addressing a refrigerated pharmaceutical routing problem. While many collaborative strategies prioritize economic and environmental benefits, our approach highlights a vital social indicator: maintaining vehicle flow equilibrium at each depot during collaboration. This ensures the stability of transport resources for all stakeholders, promoting sustainable collaborative logistics. The problem is formulated as a multi-depot vehicle routing problem with time windows (MDVRPTW). Three collaborative strategies using Clustering VRP ($C_{LU}$VRP) and improved Open VRP (OVRP) are proposed and compared. We develop two approaches to address traditional OVRP limitations in ensuring vehicle flow equilibrium at each depot. Our models consider perishable pharmaceuticals and time-dependent travel speeds. Three hybrid heuristics based on Simulated Annealing and Variable Neighborhood Search (SAVNS) are proposed and evaluated for efficacy. Computational experiments and a case study demonstrate distinct sustainable benefits across various strategies, offering valuable insights for decision-makers in the refrigerated logistics market.

## 1. Introduction

The global refrigerated logistics market has been thriving in recent years, with a projected Compound Annual Growth Rate (CAGR) of 7.2%. It is expected to reach a total market value of USD 198.35 billion by 2030 (Polaris Market Research & Consulting 2023). Meanwhile, the pharmaceutical industry has also experienced tremendous growth over the last decade due to an increasing demand for pharmaceutical distribution. The pharmaceutical refrigerated logistics market has captured a revenue share of over 28.0% in 2021, and is expected to expand at a significant CAGR of 7.4% from 2022 to 2030 (Statista 2021). Generally, pharmaceutical products need to be stored at a precise temperature to maintain their efficiency. Transporting cold-chain drugs like vaccines and insulin should be kept at low temperatures, usually between 2°C and 8°C. Therefore, it is critical to transport pharmaceuticals by refrigerated vehicles within limited time. However, refrigerated vehicles constitute a pivotal component in the financial investment of distributors (Chu et al., 2006). This is primarily due to their elevated operational costs, which subsequently prompt individual distribution centers to limit the number of delivery vehicles at their disposal (Wang et al. 2018). Consequently, this practice leads to a lower utilization efficiency of the finite transport resources, particularly when catering to the demanding logistics of perishable item deliveries. Additionally, refrigerated logistics have led to a significant amount of carbon emissions due to energy consumption and good losses (Ghadge et al. 2020). As a result, much attention has been given to the study of sustainable refrigerated logistics, triggered by changes in government policies, climate change, and consumer pressure (Mahapatra et al. 2021). Public institutions are encouraging businesses to collaborate to improve sustainability and efficiency. To achieve this, horizontal collaboration serves as an efficient solution, especially in the transportation and logistics sector, creating win-win situations for companies operating at the same level of the supply chain (Molenbruch et al. 2017). However, various collaborative strategies exhibit different efficiencies in emissions reduction and cost reduction (Benjaafar et al. 2013). Therefore, companies should reconsider their collaborative strategies, and all players should focus not only on reducing operational costs but also on enhancing sustainability (Prajapati et al. 2022).

Meanwhile, the triple bottom line (TBL) of sustainability, which encompasses economic, environmental, and social dimensions (Hosseini-Motlagh et al. 2021) is gaining momentum in recent years. However, the majority of existing research on refrigerated logistics only focuses on the first two pillars, neglecting the social dimension (Mogale et al. 2020). Mahdinia et al. (2018) emphasized the importance of social impacts when developing sustainable transportation networks. According to the Global Reporting Initiative (GRI), the social dimension refers to the organization impacts on the social systems within which it operates (2013). Specifically, social issues like balancing challenges are vital components of the social dimension but receive less attention in sustainable refrigerated logistics research (Kumar et al. 2022). To fill this gap, we use the pharmaceutical industry as an example and propose a collaborative refrigerated distribution network integrating all dimensions of TBL. Our models significantly emphasize achieving vehicle flow equilibrium at each depot, ensuring stable transport resources for all stakeholders, and promoting consistent performance in long-term collaborative logistics. As an illustration, the delivery of pharmaceuticals (e. g. vaccines, insulin, blood etc.) has always been a race against time. Inefficient deliveries may contribute to increased drug losses, resulting in elevated CO2 emissions and financial expenses, diminished customer satisfaction, and potentially jeopardizing the well-being of patients. Simultaneously, in collaboration, distribution centers may possess varying transport resources influenced by factors such as operational costs and geographical locations, leading to different numbers of available refrigerated vehicles. Therefore, ensuring the stability of transport resources for the distributors is crucial in fostering sustainable collaboration. In this regard, the balance of vehicle flow for each distributor becomes a key indicator of sustainable logistics. Hence, this study proposes sustainable collaborative strategies to develop routing plans that are efficient and meet the mentioned requirements. The proposed strategies not only apply to pharmaceutical industry, but also to other categories of refrigerated logistics, such as agricultural, meat, and other products.

The problem under study can be viewed as a multi-depot vehicle routing problem with time windows (MDVRPTW). Two state-of-the-art strategies commonly applied to MDVRPTW are Clustering VRP ($C_{LU}VRP$) and Open VRP (OVRP). In $C_{LU}VRP$, individual customers are partitioned into predefined clusters, and each customer is assigned to a specific depot (Defryn and Sorensen 2018). On the other hand, OVRP enables distributors to collaborate and share transport resources, allowing vehicles to depart from and return to different depots, thereby enhancing logistics efficiency and reducing more distribution costs compared to $C_{LU}VRP$ (Fleszar et al. 2009). Bao and Zhang (2018) integrated green logistics with route optimization of cold chain logistics using the OVRP strategy and demonstrated that OVRP outperforms $C_{LU}$VRP with respect to transportation costs and CO2 emission costs.

Previous studies indicate that OVRP outperforms $C_{LU}$VRP in terms of economic and environmental aspects during distribution. However, most of them disregard the social dimension of collaboration, particularly the vehicle flow equilibrium for each stakeholder, which directly impacts the stability of transport resources. As to this, we observe that $C_{LU}$VRP maintains the vehicle flow equilibrium consistently, whereas OVRP has a limitation that could result in potentially imbalanced vehicle flow to the depots after distribution. This imbalance may impact collaboration in subsequent rounds, reducing its sustainability. Therefore, our goal is to investigate whether improvements to OVRP still provide economic and environmental benefits over $C_{LU}$VRP. Additionally, we aim to shed light on the social impacts of these collaborative strategies and determine their sustainable potential. The main contributions of the study are summarized below:

(1) We introduce a sustainable routing problem in pharmaceutical refrigerated logistics. Three collaborative strategies based on the $C_{LU}$VRP and improved OVRP are proposed. Specifically, we focus on achieving vehicle flow equilibrium at each depot during collaboration to ensure the stability of transport resources for the depots.

(2) We develop mixed-integer linear programming (MILP) models to address the unique challenges posed by time-dependent travel speeds and perishable goods in refrigerated logistics. Our models accurately reflect CO2 emissions and travel times in real-world scenarios.

(3) We propose hybrid heuristics that use Simulated Annealing and Variable Neighborhood Search (SAVNS) algorithms tailored to our collaborative strategies. We perform comparative experiments to validate their effectiveness and rationality.

(4) We conduct computational experiments and sensitivity analyses to compare the sustainable performance of the strategies. A case study on vaccine distribution and extended analysis are carried out to demonstrate the benefits of collaboration and provide managerial insights for decision-making based on the characteristics of different strategies.

The remainder of the paper is organized as follows. Section 2 reviews the related literature. Section 3 presents the problem and the corresponding mathematical models. Section 4 provides a detailed description of the proposed hybrid heuristics, and the results of extensive computational experiments are presented in Section 5. Finally, Section 6 contains concluding remarks and managerial insights for decision makers in collaborative refrigerated logistics.

**2. Literature review**

The problem under investigation concerns the vehicle routing problem (VRP) in the context of sustainable refrigerated logistics, particularly in pharmaceutical distribution. Therefore, this section will review relevant literature in two streams: (1) VRP studies on sustainable refrigerated logistics, and (2) related VRP variants, including sustainable VRP and collaborative VRP.

*2.1 VRP studies on sustainable refrigerated logistics*

Refrigerated logistics have led to increasing energy consumption due to the rising number of circulating refrigerated vehicles (Maiorino et al. 2021). It is estimated that global refrigerated logistics accounts for approximately 15% of fossil fuel energy consumption (Adekomaya et al. 2016). In recent years, numerous studies have been conducted to promote the sustainability of refrigerated logistics through vehicle routing optimization (Maiorino et al. 2021). Wang et al. (2017) investigated a sustainable VRP with time windows (VRPTW) for refrigerated logistics based on carbon tax in China. Their objective function for the lowest cost included refrigeration costs, penalty costs, and carbon emission costs. Zhao et al. (2020) optimized logistics cost, carbon emissions, and customer satisfaction in a refrigerated logistics routing problem. Awad et al. (2020) conducted a review of VRP solutions implemented in refrigerated logistics, finding that the general objective function was to minimize the total distribution cost related to transportation, product quality, and environmental factors/constraints. Meanwhile, supply chain collaboration has been studied to make overall improvements in reducing costs and emissions (Töptal et al. 2014). Wang et al. (2021) applied $C_{LU}$VRP with an extended k-means clustering to study collaborative refrigerated logistics with resource sharing and temperature control constraints. Shi et al. (2022) designed a refrigerated logistics scheduling system considering economic, safety, and environmental factors. They addressed an OVRP with time windows to achieve resource coordination. Research on the sustainability of the refrigerated supply chain has drawn increased attention from both researchers and industrial practitioners in recent years. However, this area remains underexplored (Narwaria 2018). Kumar et al. (2022) pointed out that many supply chain players hold misconceptions that sustainability only refers to the environment, lacking a full understanding of the TBL.

Pharmaceutical distribution is an essential aspect of refrigerated logistics. The operational management literature on this topic mostly deals with VRP models that have different objectives. For instance, early research conducted by Moghadam and Seyedhosseini (2010) achieved the optimization of drug distribution routes by minimizing the total cost of refrigerated storage, transportation, and fixed vehicle costs. In the context of epidemics, Won and Lee (2020) stressed the need for knowledge sharing and collaboration to discover novel solutions to refrigerated logistics challenges in potential future virus outbreaks. Janatyan et al. (2021) proposed an optimizing model for sustainable pharmaceutical distribution, encompassing all aspects of the TBL. However, their study only addressed the distribution of non-cold medicines, and the sustainable concerns related to refrigeration were not within the scope of their research. The authors considered the minimal cost, minimal CO2 emissions, and maximum job creation as the economic, environmental and social aspects, respectively. The definitions of the social dimension in TBL are diverse, covering a wide range of social factors. Mota et al. (2015) suggested a social indicator suited for evaluating strategic decisions by considering the influence of social and political matters on the performance of companies. Ding (2018) reported there was a lack of discussion on traditional social issues in the sustainable pharmaceutical supply chain research, such as appropriate work patterns, relationships between organizations and individuals, balancing concerns, etc. He pointed out that ineffective collaborations can hinder the inclusion of sustainability.

*2.2 Related VRP variants: sustainable VRP and collaborative VRP*

Driven by the need for a more holistic and sustainable approach to logistics optimization, scholars have begun to expand the optimization goal by incorporating environmental and social aspects, in addition to solely considering economic benefits. This is known as sustainable VRP (Sbihi and Eglese 2010). In terms of environmental impacts, although there are ontological differences between 'sustainability' and 'green', they are used interchangeably in the VRP literature (Dündar et al. 2021). We direct readers to the literature review on green VRP (GVRP) by Moghdani et al. (2021), which included variants of

GVRP, objective functions, and solutions approaches. Dündar et al. (2021) reviewed sustainable VRP studies and found that those considering social dimensions were based on people, product, and time. Time-based factors are most commonly used in social dimensions. Soysal et al. (2018) used delayed delivery minimization in their studies. Scaburi et al. (2020) conducted a case study on vehicle routing with full considerations of the TBL. They took operational costs, carbon emission, and customer satisfaction as the economic, environmental, and social indicators, respectively. Abdullahi et al. (2021) investigated sustainable VRP research integrating the three dimensions of the TBL. They further explored the trade-off between these sustainability dimensions, measured by vehicle distribution costs, $CO_2$ emissions costs, and accident risk costs. The authors pointed out that more research is needed to consider social dimensions such as balancing concerns in sustainable VRP.

In collaborative VRP, participants in the distribution network can share vehicles to optimize the use of their transport resources and avoid empty truck transportation (Dai and Chen 2012). Horizontal collaboration in logistics management leads to the fulfillment of not only economic but also ecological purposes, such as reducing road congestion and emissions of hazardous substances (Gansterer and Hartl 2018). Many collaborative VRP studies have accounted for sustainable aspects. Sanchez et al. (2016) proposed a VRPTW in a collaborative setting, where different companies pool resources to reduce carbon emissions. They demonstrated that their model can cut the carbon emissions by 60% and achieve nearly 55% cost savings. Wang et al. (2018) optimized urban logistics by simultaneously optimizing logistics costs and transportation carbon emissions, focusing on horizontal collaboration and equitable benefit sharing in MDVRP. It is worth noting that both economic and environmental dimensions have been extensively studied in sustainable VRP and collaborative VRP, while social dimensions have received less attention (Abdullahi et al. 2021). As to this, Reyes-Rubiano et al. (2020) proposed a metaheuristic-based approach for solving an enriched MDVRP, taking TBL into account and considering the risk of accidents from the social viewpoint. However, the social dimension comprises subjective and complex components that can be based on customer-employee perspectives to propose specific indicators (Delucchi and McCubbin 2010).

In the context of long-term collaborative distribution, maintaining stable transport resources for each distributor is of utmost importance. This directly influences the logistic efficiency of every distributor in each round, as it determines the number of routes required to fulfill the demand (Fikar 2018). Several studies have considered the transport resources. Hariga et al. (2017) considered the limited transport resources of a depot, treating the number of vehicles in the optimization model as a decision variable rather than a constraint. Babagolzadeh et al. (2019) developed a two-echelon OVRP model with time windows, considering the influence of available transport resources and work load of each intermediate depot. Nevertheless, while achieving stable transport resources for each distributor can be accomplished by vehicle flow equilibrium at each depot, we notice that studies focusing on vehicle flow equilibrium are less common in the MDVRPTW field. In contrast, this aspect finds greater recognition and attention within the context of sustainable management for bike/vehicle sharing schemes, particularly when addressing challenges related to re-distribution or re-balancing issues. For example, Li and Liu (2021) investigated a static bike re-balancing problem with optimal user incentives, aiming to minimize the bike flow balancing cost, including travel costs, imbalanced penalties, and incentive costs. Several related studies exist on this topic, and interested readers can refer to a comprehensive review conducted by Illgen and Höck (2019). This review encompasses modeling efforts pertaining to vehicle relocation in vehicle sharing systems, all striving to achieve equilibrium in vehicle flow at each location.

To the best of our knowledge, no studies have addressed the social issue of maintaining the stable transport resources for each stakeholder by achieving vehicle flow equilibrium in sustainable or collaborative MDVRPTWs, nor in studies related to refrigerated logistics routing. In this regard, our study takes an operational lens and examines the differences in sustainable performance among various collaborative strategies. Regarding the solution approach, while MDVRPTWs typically apply to large-scale instances, and heuristics are effective solution methods for problems of this kind, none of the studies discussed in the literature explored the potential of hybrid heuristics based on Simulated Annealing and Variable Neighborhood Search. Table 1 provides a list of literature related to our work.

Table 1. Related papers classification

| Reference | The Triple Bottom Line (TBL) | | | | | | | | | Resolution |
|---|---|---|---|---|---|---|---|---|---|---|
| | Economic | | | | Environmental | Social | | | | |
| | | | | | | | | Collaboration | | |
| | Op | Dis | GL | Ref | CE | CS | BC | $C_{LU}VRP$ | OVRP | |
| Wang et al. (2017) | √ | √ | √ | √ | √ | | | | | Evolutionary Genetic Algorithm |
| Zhao et al. (2020) | √ | √ | √ | | √ | | | | | Ant Colony Algorithm |
| Wang et al. (2021) | √ | √ | | | | | | √ | | Tabu search |
| Shi et al. (2022) | √ | | | | √ | √ | | | √ | NSGA-II |
| Moghadam and Seyedhosseini (2010) | √ | √ | | √ | | | | | | Particle Swarm Optimization |
| Janatyan et al. (2021) | | √ | | | √ | | √ | | | NSGA-II and Particle Swarm Optimization |
| Mota et al. (2015) | √ | √ | | | √ | √ | | | | Metaheuristic |
| Xiao and Konak (2015) | | √ | | | √ | | | | | Simulated Annealing |
| Soysal et al. (2018) | √ | √ | | | √ | | | | | CPLEX solver |
| Scaburi et al. (2020) | | √ | | | √ | | | | | LINGO software |
| Abdullahi et al. (2021) | √ | √ | | | √ | | | | | Biased-randomised Iterated Greedy Algorithm |
| Sanchez et al. (2016) | √ | √ | | | √ | √ | | | | Scatter Search Metaheuristic |
| Wang et al. (2018) | | √ | | | | √ | √ | | | Improved NSGA-II |
| Reyes-Rubiano et al. (2020) | √ | √ | | | √ | | | | | Biased-randomized Variable Neighborhood Search |
| Hariga et al. (2017) | | √ | √ | √ | √ | | | | | Iterative Search |
| Babagolzadeh et al. (2019) | √ | √ | | | √ | | | | √ | - |
| Li and Liu (2021) | | √ | | | | | | | | Bi-level Variable Neighborhood Search |
| **The current paper** | **√** | **√** | **√** | **√** | **√** | **√** | **√** | **√** | **√** | **Simulated Annealing-Variable Neighborhood Search** |

Op: Operation, Dis: Distance, GL: Good Loss, Ref: Refrigeration, CE: CO$_2$ Emission, CS: Customer Satisfaction, BC: Balancing Concern

## 3. Problem description and formulation

### 3.1 Problem description

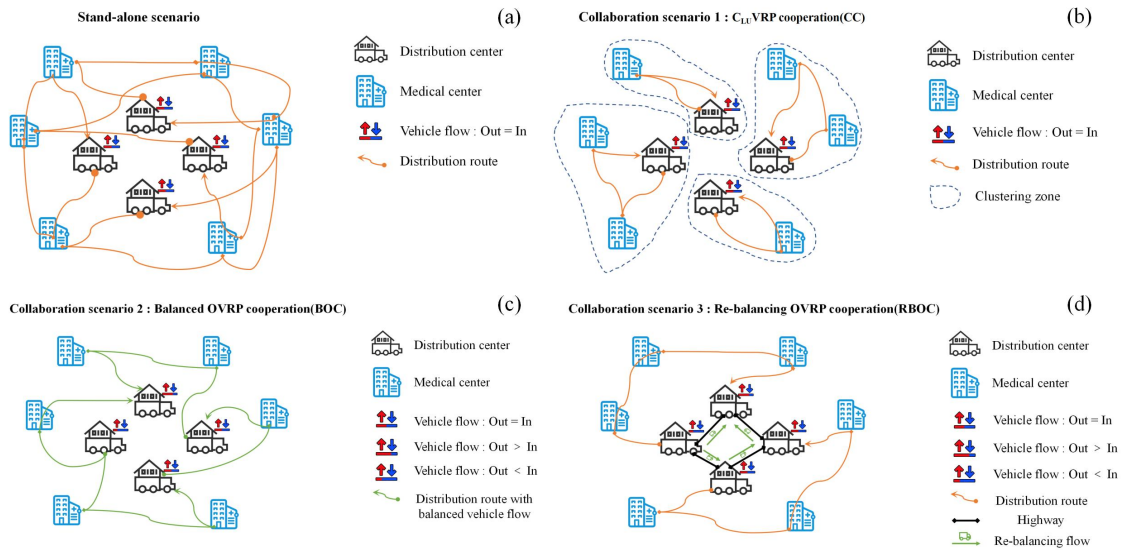

Figure 1 Caption: (a)-(d) Different scenarios in pharmaceutical refrigerated logistics

Figure 1 Alt Text: Illustrations of the four distribution scenarios: (a) Stand-alone scenario; (b) CLUVRP cooperation (CC); (c) Balanced OVRP cooperation (BOC); (d) Re-balancing OVRP cooperation (RBOC).

The understudied MDVRPTW addresses the distribution of a single type of refrigerated drug from several distribution centers to multiple medical centers. We denote the set of distribution centers as $M$ and the set of medical centers as $N$. We assume that distribution centers are uniformly dispersed among the medical centers in a metropolitan area, resulting in a relatively balanced number of nearest medical centers around each distribution center. Homogeneous vehicles are utilized for the delivery, which have time-dependent travel speeds. Each medical center has a soft time window restriction for distribution. In a stand-alone scenario, a medical center can receive separate orders from different distribution centers (as shown in Figure 1. (a)), but distribution centers can collaborate horizontally to better utilize limited transport resources and save costs. In collaboration scenarios, the aggregate of orders constitutes the demand of each medical center $I_n \in N$, which can only be visited once by a distribution center $D_m \in M$. We assume that the locations of all facilities, and the limited transport resources of each distribution center are known. This research focuses on evaluating different collaborative strategies for routing solutions in terms of sustainability. In light of the limited focus on transport resource stability through vehicle flow equilibrium at each depot in most horizontal collaborations, we address this gap by introducing three strategies, as shown in Figure 1.(b)-(d):

- $C_{LU}$VRP cooperation (CC): Figure 1.(b) shows that in the CC strategy, medical centers are clustered before being assigned to a particular depot. It ensures vehicle flow equilibrium as in classical VRP where the vehicle departs from and return to the same depot.

- Balanced OVRP cooperation (BOC): the BOC strategy proposes an improved OVRP mode, as shown in Figure 1.(c). The vehicles are allowed to return to a different depot, but the vehicle flow for each depot can be balanced synchronously during distribution.

- Re-balancing OVRP cooperation (RBOC): the RBOC strategy is inspired by traffic control measures introduced by governments to alleviate traffic congestion and environmental pollution (Yağmur and Kesen 2022). It facilitates depot collaboration in the traditional OVRP mode while enabling cost-effective vehicle transfers between depots through government subsidies. This achieves vehicle flow equilibrium after distribution (see Figure 1.(d)).

More implementation details for the three collaborative strategies are supplied in Section 4.

*3.2 Mathematical formulation*

To address the aforementioned issues, we develop a basic mixed-integer linear programming model (MILP) that incorporates features of the MDVRPTW and the cold-chain drug distribution network. The components of this model are applicable to all the collaborative models. We assume that each vehicle starts off at 9:00 AM and must complete delivery within 8 hours (9:00 AM-17:00 PM), and that variations in traffic congestion throughout the day will affect the average travel speed of the vehicles, i.e., time-dependent travel speeds. Unlike models proposed by related works, most of which assumed that the vehicle's speed remains constant on the same arc (Zhou et al. 2022), we consider the impact of client service time on vehicle speed variations, as well as the possibility that the vehicle's speed may undergo multiple changes on the same arc. This adds complexity to the computation of vehicular travel time, which involves the calculation of the objective function presented below. Moreover, the model involves multiple depots. To clarify the departure and return depots for each route, we introduce a replica set $M^*$ as the return depots set, where each depot in $M^*$ corresponds to the same depot in $M$.

The indices and parameters are presented below:

| | Sets and indices |
|---|---|
| $M$ | Set of distribution centers, $\{D_1, D_2, D_3, \ldots, D_m\}$ |
| $M^*$ | Replica set of distribution centers, $\{D_1^*, D_2^*, D_3^*, \ldots, D_m^*\}$ |
| $N$ | Set of medical centers, $\{I_1, I_2, I_3, \ldots, I_n\}$ |
| $PN$ | Set of physical nodes in the network, $PN = M \cup N$ |
| $PN^*$ | Set of physical nodes in the network, $PN^* = M^* \cup N$ |
| $L$ | Set of arcs in the network |
| $\Pi$ | Set of time periods, $\{\pi_1, \pi_2, \pi_3, \ldots, \pi_k\}$ |
| $H$ | Set of vehicles |
| | **Parameters** |
| $\omega_i$ | Total demand for cold-chain drugs by medical center $i \in N$ |
| $Q_h$ | Maximum load capacity of vehicle $h \in H$ |
| $G_i$ | Number of refrigerated vehicles owned by distribution center $i \in M$ |
| $f$ | Fixed cost of operating a vehicle |
| $d_{ij}$ | Euclidean distance on arc $<i,j> \in L$ |
| $d_{ij}^{st}$ | Shortest travel distance between two distribution centers $i \in M$ and $j \in M$ |
| $v_\pi$ | Average travel speed of a vehicle at time period $\pi \in \Pi$ |
| $t_{ijh}^\pi$ | Travel time of vehicle $h \in H$ on arc $<i,j> \in L$ at time period $\pi \in \Pi$ |
| $t_{ijh}$ | Total travel time of vehicle $h \in H$ on arc $<i,j> \in L$ |
| $et_i$ | The earliest service start time of medical center $i \in N$ |
| $lt_i$ | The latest service start time of medical center $i \in N$ |
| $c_0$ | Coefficient for travel cost during distribution |
| $\alpha$ | Discount of travel unit cost after distribution |
| $c_1$ | Coefficient for refrigeration cost |
| $c_2$ | Coefficient for penalty cost of too early arrival |
| $c_3$ | Coefficient for penalty cost of too late arrival |
| $\beta$ | Coefficient for good loss in distribution/unloading phase |
| $\lambda$ | Coefficient for carbon emission |
| $e$ | Coefficient for carbon cost |
| $E_{ih}^{\omega_i}$ | CO2-emission cost of vehicle $h \in H$ serving node $i \in PN$ with $\omega_i$ drug demand |
| $TT_i$ | Service time required at medical center $i \in N$ |
| $u_i$ | Difference between the number of vehicles returning to and departing from depot $i \in M$ after distribution |
| $W$ | A large number |
| | **Decision variables** |
| $X_{ijh}$ | 1 if vehicle $h$ travels on arc $<i,j> \in L$, 0 otherwise |
| $A_{ih}$ | Arrival time of vehicle $h \in H$ at node $i \in PN$ |
| $Y_{ih}^\pi$ | 1 if vehicle $h \in H$ departs from node $i \in PN$ at time period $\pi \in \Pi$, 0 otherwise |
| $Z_{ij}$ | Number of vehicles transported from distribution center $i$ to $j$, $i, j \in M$ |

**Basic Model**

**Objective function**

$$min \sum_{i \in M, j \in N} \sum_{h \in H} f X_{ijh} + \sum_{i \in PN, j \in PN^*} \sum_{h \in H} c_0 d_{ij} X_{ijh} + \sum_{i \in PN, j \in PN^*} \sum_{h \in H} E_{jh}^{\omega_j} X_{ijh} + \\ \sum_{i \in PN, j \in N} \sum_{h \in H} (c_1 + \beta \omega_j)(t_{ijh} + TT_j) X_{ijh} + \sum_{\pi \in \Pi} \sum_{i \in N} \sum_{h \in H} Y_{ih}^\pi (c_2(et_i - A_{ih})^+ + c_3(A_{ih} - lt_i)^+) \quad (1)$$

Our model aims to determine an efficient routing plan that minimizes the total costs, whether in stand-alone or collaboration scenarios. The objective function (1) minimizes various costs considering sustainability, which includes: fix cost, transportation cost, CO2-emission cost, cooling cost, good loss cost due to distribution/unloading phase, and penalty cost for time window violations. The fix cost refers to the cost generated by utilizing a vehicle in each route. Both the cooling cost and good loss cost are set to be proportional to spending time during distribution. We address the non-linear penalty cost in the objective function by setting any negative values to zero and considering only positive values. To account for this issue, we introduce auxiliary variables (see Appendix 1 for specific modelling).

We will now provide a detailed account of how to measure CO2 emissions of a visiting node and the travel time on any arc, which are intricately linked to time-dependent travel speeds and therefore pose challenges to the objective calculation (see the illustration of a delivery route in Figure 2).

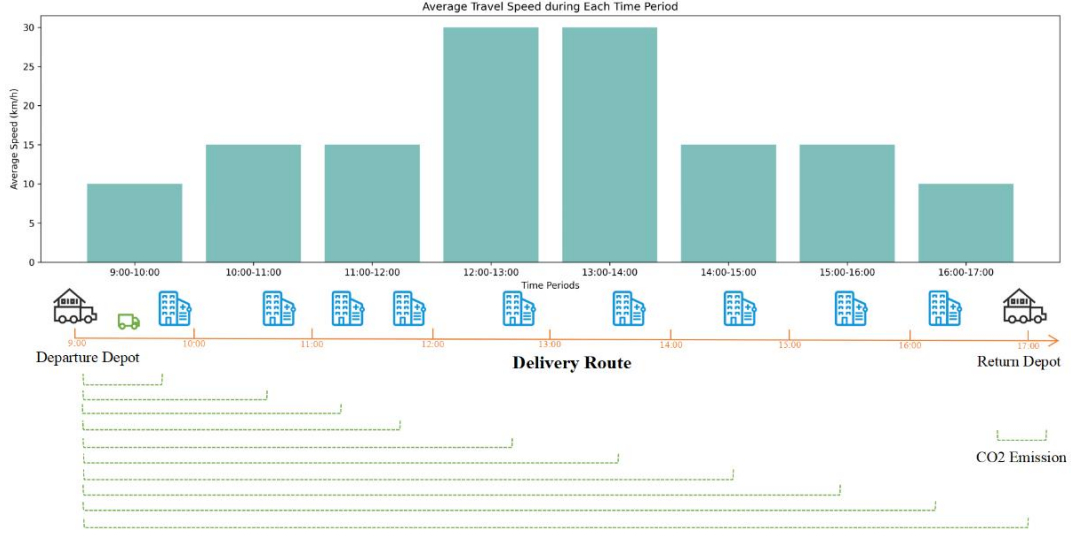

Figure 2 Caption: Illustration of a delivery route with time-dependent travel speeds

Figure 2 Alt Text: The top graph depicts the vehicle's speed over time, while the middle section illustrates its customer visits starting at 9 A.M. The green lines below represent CO2 emissions during these visits.

*Measurement of CO2 emissions.* CO2 emissions are determined by distance travelled and loading of the vehicles (Messaoud et al. 2018). $E_{ih}^{\omega_i}$ represents the cost of CO2 emission by vehicle $h \in H$ serving node $i \in N$ with demand $\omega_i$. To calculate the distance vehicle $h$ travelled until arriving at $i$, let arc set $\theta_{ih}$ represent the sub-string route $h$ travelled before visiting $i$, where each node $j$ ($j \neq i$) contained in $\theta_{ih}$ has an earlier arrival time than $i$ ($A_{jh} \leq A_{ih}$). Then, $E_{ih}^{\omega_i}$ can be given by:

$$E_{ih}^{\omega_i} = e\lambda(P_0 + \omega_i(P^* - P_0)/Q_h) \sum_{<j_1,j_2> \in \theta_{ih}} d_{j_1 j_2} \tag{2}$$

Where $P_0$ (resp. $P^*$) refers to the fuel consumption per unit distance when the vehicle is unloaded (resp. full loaded), and $e$ (resp. $\lambda$) refers to the coefficient for carbon cost (resp. carbon emission).

*Measurement of travel time on arc* $<i,j> \in L$. The cooling cost and good loss cost are closely related to the travel time $t_{ijh}$ of vehicle $h$ on arc $<i,j>$. As Figure 2 shows, we divide a day into equally lengthy time periods, and the average travel speed of a vehicle during distribution is a function segmented into time periods (Ichoua et al. 2003). $t_{ijh}$ can be calculated in the MILP with two steps and the corresponding heuristic algorithm is proposed in the following pseudo-code:

```
Input:  Time periods set Π = {π₁,π₂,π₃,...,π_k}; each π in Π has a start
        moment bt_π, end moment et_π, and average travel speed v_π;
        dep_{ih} ← (A_{ih} + TT_i) as the departure moment of vehicle h from node i;
        d_{ij} as the Euclidean distance between node i and j;
Output: t_{ijh} /* Travel time of vehicle h on arc <i,j> */;
1  for each time period π do
2    |  if bt_π ≤ dep_{ih} ≤ et_π then
3    |  |  π_s ← π as the departure time period of vehicle h from node i;
4    |  |  Set the speed of the vehicle h as v_{π_s};
5    |  end
6  end
7  resd ← d_{ij} /* Define the rest of distance untravelled */;
8  t_{ijh} ← 0;
9  tempt ← dep_{ih};
10 p ← s;
11 while resd! = 0 do
12   |  tempd ← (et_{π_p} − tempt)×v_{π_p};
13   |  if tempd ≤ resd then
14   |  |  t_{ijh} ← t_{ijh} + et_{π_p} − tempt;
15   |  |  resd ← resd − tempd;
16   |  else
17   |  |  t_{ijh} = t_{ijh} + resd/v_{π_p};
18   |  |  break;
19   |  end
20   |  p = min(p + 1, k);
21   |  tempt = bt_{π_p};
22 end
23 return t_{ijh}
```

(1) Noting that each time period $\pi \in \Pi$ spans from $bt_\pi$ to $et_\pi$ with an average travel speed $v_\pi$, we can determine the time period $\pi_s$ at which the vehicle $h$ departs from node $i$ ($Y_{ih}^{\pi_s} = 1$) by constraints (3)-(5), and its corresponding travel speed is denoted as $v_{\pi_s}$:

$$Y_{ih}^\pi \leq \sum_{j \in PN^*} X_{ijh} \, max(bt_\pi - A_{ih} - TT_i, 0) \quad \forall i \in PN, \forall h \in H, \forall \pi \in \Pi \tag{3}$$

$$Y_{ih}^\pi \leq (1 - W) \sum_{j \in PN^*} X_{ijh} \, min(A_{ih} + TT_i - et_\pi, 0) \quad \forall i \in PN, \forall h \in H, \forall \pi \in \Pi \tag{4}$$

$$\sum_{\pi \in \Pi} Y_{ih}^\pi = \sum_{j \in PN^*} X_{ijh} \quad \forall i \in PN, \forall h \in H \tag{5}$$

Constraints (5) indicates that only when the vehicle $h$ visits node $i$ do we discuss the value of $Y_{ih}^\pi$.

(2) We denote $\partial_{\pi_s} = \{\pi_s, \pi_{s+1}, \pi_{s+2}, \ldots, \pi_k\}$ as the set of time periods from $\pi_s$ on forwards till the last time period $\pi_k$ in the planning horizon. The travel time ($t_{ijh}^{\pi_p}$) of vehicle $h$ on arc $<i,j>$ at time period $\pi_p \in \partial_{\pi_s}$, and the total travel time ($t_{ijh}$) of $h$ on arc $<i,j>$ can be obtained by formulas (6)-(7):

$$t_{ijh}^{\pi_p} = \begin{cases} min(et_{\pi_p} - A_{ih} - TT_i, \dfrac{d_{ij}}{v_{\pi_p}}) & p = s \\ min(\dfrac{d_{ij} - \sum_{n=s}^{p-1}(v_{\pi_n} t_{ijh}^{\pi_n})}{v_{\pi_p}}, et_{\pi_p} - bt_{\pi_p}) & \text{otherwise} \end{cases} \tag{6}$$

$$t_{ijh} = \sum_{\pi_p \in \partial_{\pi_s}} t_{ijh}^{\pi_p} \tag{7}$$

Other constraints are:

$$\sum_{j \in N} X_{ijh} \leq 1 \quad \forall i \in M, \forall h \in H \tag{8}$$

$$\sum_{j \in N} X_{jih} \leq 1 \quad \forall i \in M^*, \forall h \in H \tag{9}$$

$$\sum_{j \in PN^*} X_{ijh} = \sum_{j \in PN} X_{jih} \quad \forall i \in N, \forall h \in H \tag{10}$$

$$\sum_{h \in H} \sum_{j \in PN^*} X_{ijh} = 1 \quad \forall i \in N \tag{11}$$

$$\sum_{h \in H} \sum_{j \in PN^*} X_{ijh} \leq G_i \quad \forall i \in M \tag{12}$$

$$\sum_{i \in N} \sum_{j \in PN^*} \omega_i X_{ijh} \leq Q_h \quad \forall h \in H \tag{13}$$

$$A_{ih} + TT_i + X_{ijh} t_{ijh} - W(1 - X_{ijh}) \leq A_{jh} \quad i \in PN, j \in PN^*, \forall h \in H \tag{14}$$

$$X_{ijh} \in 0,1 \quad \forall i,j \in PN \cup PN^*, \forall h \in H \tag{15}$$

$$Y_{ih}^\pi \in 0,1 \quad \forall i \in N, \forall h \in H, \forall \pi \in \Pi \tag{16}$$

$$A_{ih} \geq 0 \quad \forall i \in PN, \forall h \in H \tag{17}$$

Constraints (8)-(9) are depot constraints ensuring that each vehicle starts and ends its route at a certain depot. Constraints (10)-(11) ensure that the route remains continuous and the demand of each customer is satisfied without revisiting the same customer. Constraint (12) indicates that the number of vehicles departing from each depot does not exceed the available transport resources at that depot. Constraint (13) represents that the number of drugs allocated to the vehicle cannot exceed the maximum load capacity. Constraint (14) indicates the time relationship between the antecedent and successor nodes to eliminate the sub-tour in the network. Lastly, Constraints (15) - (17) define feasible values for decision variables.

To apply the basic model (1)-(17) to stand-alone distribution, we add constraint (18) to ensure the vehicle returns to the initial depot at the end of its route, noting that each depot $i \in M$ has a corresponding depot $i^* \in M^*$:

$$\sum_{j \in N} X_{ijh} = \sum_{j \in N} X_{ji^*h} \quad \forall i \in M, \forall i^* \in M^*, \forall h \in H \tag{18}$$

**Collaborative Models**

*CC Model.* The CC strategy follows a classic $C_{LU}VRP$ mode, in which medical centers are clustered beforehand and are assigned to different distribution centers. Therefore, vehicles work in the same way as in the stand-alone scenario, except for the different service coverage and allocated drug demand for each vehicle. The model (1)-(18) fits well in this strategy.

*BOC Model.* In traditional OVRP, vehicles can choose their departure and return depots freely to avoid detours, thereby having lower distribution costs. However, this might give rise to an imbalanced vehicle flow at the distribution centers. To eliminate this imbalance during distribution, we introduce constraint (19) by modifying constraint (18):

$$\sum_{h \in H} \sum_{j \in N} X_{ijh} = \sum_{h \in H} \sum_{j \in N} X_{ji^*h} \quad \forall i \in M, \forall i^* \in M^* \tag{19}$$

Constraint (19) indicates that the vehicles don't have to go back to their departure depot, while the number of vehicles departing from each depot equals the number of vehicles returning to that depot. Therefore, BOC model can be represented by (1)-(17), (19).

*RBOC Model.* The RBOC strategy, similar to the BOC strategy, also adopts an OVRP mode but deals with distribution and vehicle flow equilibrium at each depot independently. It is assumed that highways connect any distribution center with other centers within a restricted distance. Consequently, all distribution centers can create a network utilizing these existing highways, and the shortest distance $d_{ij}^{st}$ between any two depots ($i$ and $j$) can be determined by applying the Floyd algorithm (Floyd, 1962) to compute the shortest path between them. We set $\alpha$ as the reduction ratio of travel unit cost (discount) due to government subsidies based on the assumption mentioned in Section 3.1. In the later analysis, we will address the impact of $\alpha$ on the decision maker's choice of collaborative strategies. Formula (21) calculates the difference between the number of vehicles returning to and departing from depot $i$ resulting from an imbalanced vehicle flow after distribution. The re-balancing operation is described as follows:

$$min \sum_{i \in M} \sum_{j \in M} f Z_{ij} + (1-\alpha)c_0 \sum_{i \in M} \sum_{j \in M} d_{ij}^{st} Z_{ij} + \lambda e P_0 \sum_{i \in M} \sum_{j \in M} d_{ij}^{st} Z_{ij} \quad (20)$$

$$u_i = \sum_{h \in H} \sum_{j \in N} X_{ji^*h} - \sum_{h \in H} \sum_{j \in N} X_{ijh} \quad \forall i \in M, \forall i^* \in M^* \quad (21)$$

$$\sum_{j \in M} Z_{ij} - \sum_{j \in M} Z_{ji} + u_i = 0 \quad \forall i \in M \quad (22)$$

$$Z_{ij} \in Z^+ \quad \forall i \in M, \forall j \in M \quad (23)$$

Where formula (20) states that the re-balancing cost of those transferred vehicles includes the fix cost, transportation cost and the cost of CO2 emission. Constraints (22)-(23) indicate that depots can allocate/receive vehicles from other depots to balance its vehicle flow and the decision variable $Z_{ij}$ belongs to the set of non-negative integers. RBOC model incorporates (2)-(17), (21)-(23) as the constraints and its objective function is to minimize the sum of distribution cost in (1) and re-balancing cost in (20).

## 4. Solution approach

To efficiently solve the proposed collaborative models for different strategies, three hybrid heuristics based on Simulated Annealing and Variable Neighborhood Search (SAVNS) are developed to find near-optimal solutions. The core principle of these heuristics is to generate an initial feasible solution by a greedy algorithm. And then, customized SAVNS heuristics are designed for each collaborative strategy to improve the solution quality. In the upcoming sections, CC-SAVNS for CC strategy is presented in Section 4.1, while BOC-SAVNS (resp. RBOC-SAVNS) for BOC strategy (resp. RBOC strategy) is detailed in Section 4.2.

### 4.1 CC-SAVNS for CC strategy

The CC-SAVNS proposed in this study follows the assumption that a vehicle departs from and returns to a same depot as the classic $C_{LU}VRP$ (see Section 3). This naturally achieves vehicle flow equilibrium at each depot. The framework of CC-SAVNS is depicted in Figure 3. Firstly, the input parameters such as problem parameters, initial and final temperatures $T_0$ and $T^*$, the cooling rate $\delta$, the number of neighborhood search $k_{max}$ and the number of depots $|M|$ are set. Then a Clustering Approach (CA) is called to assign customers to the $|M|$ depots based on the shortest Euclidean distance rule. Next, a Push-Forward Insertion Heuristics (PFIH) is employed to generate initial routes for each of the $|M|$ depots. These routing plans form an initial solution of the CC model. To efficiently seek a near-optimal solution, a hybrid SAVNS (SAVNS-1) is developed.

Specifically, CA in Figure 3 draws inspiration from the Partitioning Around Medoids (PAM) approach (Zhu et al. 2015). It adheres to the assumption in Section 3.1, which assumes that depots are uniformly dispersed among numerous customers. CA involves a clustering strategy that assigns all customers one by one to the nearest depot based on the Euclidean distance between them.

Following the customer assignment by CA, the PFIH method, originally proposed by Solomon for single-depot VRPTWs, is adapted to generate an initial solution, denoted as $S_0$. PFIH iteratively inserts customer into a route based on a minimum insertion cost rule, and the details are referred to Tam and Ma (2008). If the residual load capacity of the vehicle is sufficient to accommodate the customer's demand, the customer is inserted into the current route. Otherwise, a new route is created, and the customer is assigned to this new route. This process continues iteratively until all customers are scheduled.

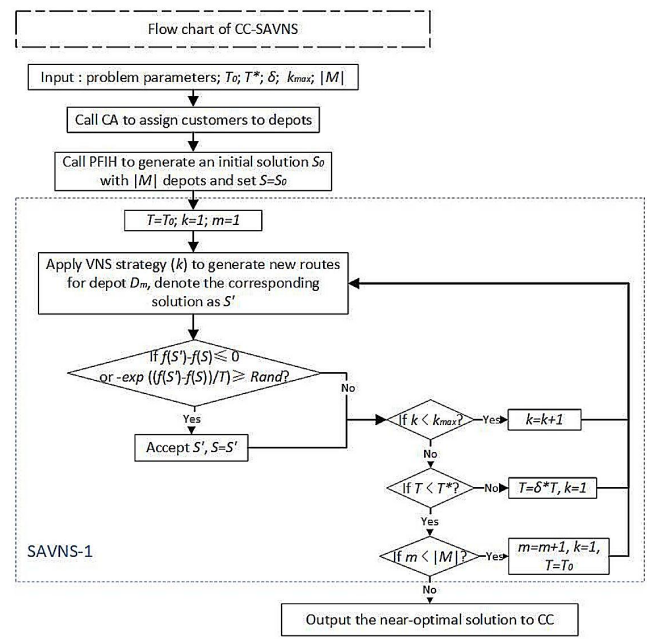

Figure 3 Caption: Flowchart of CC-SAVNS

Figure 3 Alt Text (Long Description): The CC-SAVNS involves these steps: setting input parameters, using a Clustering Approach (CA) to assign customers to depots, employing Push-Forward Insertion Heuristics (PFIH) to create initial routes for depots, and utilizing hybrid SAVNS (SAVNS-1) for near-optimal solutions.

To improve the solution quality, SAVNS-1 is developed with a main procedure consisting of three loops, as illustrated in Figure 3. Denote $S$ as the current solution, set $S = S_0$, $k=1$, $T = T_0$ and $m =1$ for depot $D_m \in M$, the first loop comprises $k_{max}$ rounds of Variable Neighborhood Search (VNS) applied to the routes of depot $D_m$ to generate its new routes, while keeping routes for other depots unchanged: After the execution of VNS strategy ($k$), the corresponding solution is denoted as $S'$. Then, the difference of objective function values $f(S') - f(S)$ between the solutions $S'$ and $S$ is calculated. If $f(S') - f(S) \leq 0$, $S'$ is accepted, if not, $S'$ is accepted with a probability according to the Metropolis criterion. And then, $k$ is updated as $k = k + 1$. The process continues until $k = k_{max}$. The second loop is related to a Simulated Annealing (SA) process. After $k_{max}$ VNS strategies are executed, the current temperature $T$ decreases to $T * \delta$. And then new $k_{max}$ VNS strategies are applied. The process repeats until $T < T^*$. Finally, the third loop is to improve the routes of each of the $|M|$ depots. Once all the depots' routes have been enhanced, the CC-SAVNS outputs the near-optimal solution. Especially, the VNS strategy ($k$) is composed of 4 steps that are detailed as follows:

**Step 1:** Randomly select two routes $r_1$ and $r_2$, and select the starting customer node $i_{r_1}$ and $i_{r_2}$ in $r_1$ and $r_2$ respectively, excluding the depot nodes of the route.

**Step 2:** From node $i_{r_1}$ (resp. $i_{r_2}$), generate a sub-path $L_{r_1}$ in $r_1$ (resp. $L_{r_2}$ in $r_2$) where the length of $L_{r_1}$ (resp. $L_{r_2}$) is the minimum length of the sub-path from node $i_{r_1}$ (resp. $i_{r_2}$) to the end of $r_1$ (resp. $r_2$) or that from node $i_{r_1}$ (resp. $i_{r_2}$) to node $i_{r_1} + k$ (resp. $i_{r_2} + k$).

**Step 3:** With a randomly generated probability, randomly apply Cross exchange or $i$-Cross exchange strategy (see Figure 4) to exchange $L_{r_1}$ and $L_{r_2}$ in $r_1$ and $r_2$.

**Step 4:** If the load capacity constraint (12) in Section 3 is satisfied, the VNS strategy ($k$) is completed, otherwise, return to Step 1.

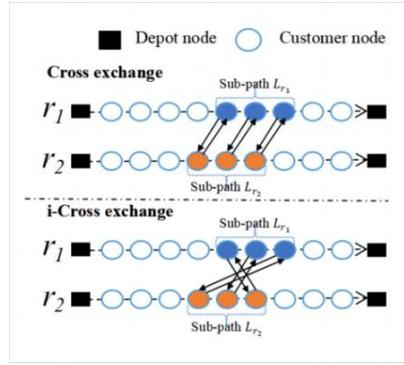

Figure 4. Cross exchange & $i$-Cross exchange strategies for $L_{r_1}$ and $L_{r_2}$ in $r_1$ and $r_2$

Figure 4 Alt Text (Long Description): $L_{r_1}$ (resp. $L_{r_2}$) is a sub-path formed from route $r_1$ (resp. $r_2$) and the length as well as starting & ending position can be different between $L_{r_1}$ and $L_{r_2}$. The upper (resp. lower) part of Figure 4 shows that Cross exchange (resp. $i$-Cross exchange) refers to exchanging the two sub-paths in forward (resp. reverse) order.

*4.2 BOC-SAVNS & RBOC-SAVNS for BOC & RBOC strategies*

BOC-SAVNS and RBOC-SAVNS introduce more flexibility in vehicle routes by relaxing the assumption that a vehicle must depart from and return to the same depot. The two algorithms share a similar framework, as illustrated in Figure 5, with the key distinction lying in their vehicle flow balancing strategies. For both algorithms, a virtual depot $D_0$ positioned at the mean coordinates of the $|M|$ depots is introduced to serve as the initial departure (resp. return) depot for all vehicles.

Differentiating from CC-SAVNS, the particularities of BOC-SAVNS (resp. RBOC-SAVNS) are as follows:

(1) PFIH in BOC-SAVNS and RBOC-SAVNS is employed to generate an initial solution with depot $D_0$, denoted as the current solution $S$. $S$ is updated by assigning the routes in $S$ to the $|M|$ depots based on the nearest depot rule, allowing for different departure and return depots for each route.

(2) In BOC-SAVNS (resp. RBOC-SAVNS), each VNS ($k$) strategy in SAVNS-2 is used to improve two randomly selected routes, And the departure and return depots of the newly updated routes are then adjusted according to the nearest depot rule in specific situations (see Figure 5). The updated solution is denoted as $S'$.

(3) In BOC (resp. RBOC) strategy, the vehicle flow equilibrium at a depot is ensured, i.e., the number of vehicles departs from a depot and return to the depot is the same, while a vehicle can depart from one depot and return to another.

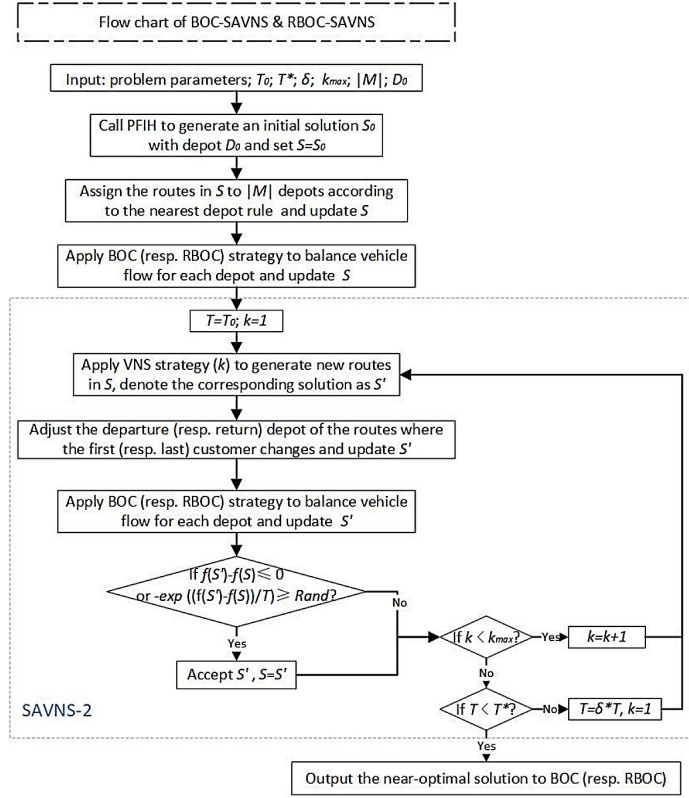

Figure 5 Caption: Flowchart of BOC-SAVNS & RBOC-SAVNS

Figure 5 Alt Text (Long Description): The BOC-SAVNS (resp. RBOC-SAVNS) involves these steps: setting input parameters, using PFIH to create an initial solution with the proposed virtual depot $D_0$, denoted as the current solution $S$, updating $S$ as $S'$ by reassigning routes to the $|M|$, further updating $S'$ with the BOC (resp. RBOC) strategy for vehicle flow equilibrium at each depot, and employing a hybrid SAVNS (SAVNS-2) to efficiently seek a near-optimal solution for the BOC (resp. RBOC) model.

The BOC strategy consists of a Balancing Approach (BA) that ensures a balanced vehicle flow at each depot during distribution, while allowing a vehicle to depart from one depot and return to another. This strategy aims to strike a favorable trade-off between vehicle flow equilibrium and transportation cost savings by permitting different departure and return depots. BA is executed in 4 steps:

**Step 1:** Calculate the difference between the number of vehicles returning to and departing from depot $D_m$ as $u_m$ ($m = 1,..., |M|$).

**Step 2:** For each route with return depot $D_m$ and $u_m > 0$, calculate the Euclidean distance $d_{im}$ between the last customer node $i$ and depot $D_m$, and store these routes in a set $R_m$ with a no creasing order of $d_{im}$.

**Step 3:** Update $S$ with the new route obtained in step 2. Set the next route in $R_m$ as the current route $r_m$. For each of the $u_m$ first routes in $R_m$, select a new nearest return depot $D_{m'}$ if $u_{m'} < 0$. And update $u_{m'} = u_{m'} + 1$.

**Step 4:** Repeat steps 2-3 until all depots $D_m$ with $u_m > 0$ are treated and the corresponding solution is denoted as $S'$.

In contrast, the RBOC strategy keeps the routes in $S$ unchanged but generates a re-balancing operation that reallocates vehicles among the depots. To achieve this, a Restricted Shortest Path Approach (RSPA) based on the Floyd algorithm is proposed, consisting of 3 steps:

**Step 1:** Calculate the difference between the number of vehicles returning to and departing from depot $D_m$ as $u_m$ ($m = 1,..., |M|$).

**Step 2:** Apply the Floyd algorithm to calculate the shortest paths between any two depots in the highway network described in the RBOC model (see Section 3.2). Note that highways only exist between depots located within a restricted distance from each other, forming a connected network encompassing all depots.

**Step 3:** Solve the linear programming model (19)-(22) using an off-the-shelf solver (e.g., Gurobi, Cplex) to generate an optimal re-balancing operation solution with the minimum cost. This solution strategically distributes the extra vehicles at each depot $D_m$ with $u_m > 0$ to each depot $D_{m'}$ with $u_{m'} < 0$.

Therefore, RSPA outputs the obtained re-balancing operation solution, which is integrated into $S$ and we denote the corresponding solution as $S'$.

## 5. Computational experiments

In this section, we begin by validating the effectiveness of the proposed models and algorithms, after which we evaluate the sustainability of the three collaborative strategies. Extensive computational experiments with sensitivity analyses are conducted to compare the sustainable performance of collaborative models. Finally, we present a case study on collaborative vaccine distribution based on real data from Shanghai, China. Certain parameter values have been pre-set in all of our simulated distribution scenarios, as shown in Table A1 (see Appendix 2). The proposed methods are implemented using MATLAB 2021 in a Windows 11 operating system on a PC with an Intel 2.60 GHz i7-1185G7 Processor and 16 GB of RAM.

### *5.1 The efficiency of the algorithms*

In Sections 5.1 and 5.2, we conduct computational experiments utilizing benchmark instances for the MDVRPTW problem provided by Coindreau et al. (2019). These instances can be accessed at https://chairelogistique.hec.ca/en/scientific-data/. We devise sets of instances specifically tailored to our problem scenarios and compare the performance of three existing solution methods (Gurobi, SA, and VNS) with our hybrid SAVNS heuristics across the three models. Notably, the existing solution methods are adapted to incorporate the characteristics of our problem. Since Gurobi fails to provide feasible solutions for larger instances, we focus on evaluating the convergence and performance of the proposed hybrid heuristics and their SA or VNS versions for such cases. We have conducted a total of 960 experiments on 16 instances, labeled as '*n--m*', where the number of customers (*n*) and depots (*m*) varied for each model. Each instance is tested five times using each approach, and the mean results are reported. Table 2 presents the results for the CC, BOC, and RBOC models, divided into three panels.

Within each panel, we provide detailed information on the runtime (*Runtime*), objective value (*Obj*), and relative improvement (*Rel.Imp.*) achieved by SAVNS approach compared to other approaches (e.g., xx-Gurobi, xx-SA, and xx-VNS, where 'xx' refers to the specific model). Namely, we have:

$$Rel.Imp. = \frac{Obj_{SAVNS} - Obj_{approach}}{Obj_{approach}} \times 100\%$$

Where $Obj_{approach}$ is the objective value of approach Gurobi, SA or VNS, and $Obj_{SAVNS}$ is the objective value obtained by the proposed hybrid SAVNS. Note that we set the result to '----' if Gurobi fails to give a feasible solution within the given maximum runtime in a number of instances.

As shown in Table 2, the best objective value obtained in each instance is highlighted in bold and the average performance of each approach across all instances is displayed at the bottom of each panel. It is evident that due to the NP-hard nature of our problem (Yu et al. 2008), Gurobi can only handle very small-scale instances, typically with no more than 30 nodes. The average relative improvement (*Rel.Imp.*) achieved by hybrid SAVNS compared to Gurobi is -0.96% in average across all the experiments involving only the small-scale instances, as Gurobi is not capable of handling large-scale. Although Gurobi consistently outperforms the proposed SAVNS in small-scale instances, the difference is acceptable considering the significant runtime savings provided by our algorithms. Gurobi requires approximately 20 times more runtime than other heuristics to find a best solution. This limitation underscores the suitability of heuristic approaches for real-world distribution scenarios with larger customer scales, often exceeding 100 nodes.

Furthermore, the proposed heuristics consistently outperform the other two heuristics across the three models, obtaining superior solutions with comparably short runtimes. Specifically, the hybrid SAVNS achieves substantial improvements in objective values for all instances, with average enhancements of 7.83%, 6.18%, and 8.51% (resp. 7.70%, 8.90%, and 8.99%) for the CC, BOC, and RBOC models, respectively, when compared to SA (resp. VNS). Notably, our algorithms exhibit enhanced performance with larger instance sizes, as evidenced by the convergence results for the '240-4' instance in Figure 6. These outcomes highlight the superior solution efficiency of hybrid SAVNS for all tested collaborative models.

Table 2. A comparision among three approaches across three models

| Instances | Runtime(s) | | | | Obj | | | | Rel. Imp. | | |
|---|---|---|---|---|---|---|---|---|---|---|---|
| (n--m) | CC-gurobi | CC-SA | CC-VNS | CC-SAVNS | CC-gurobi | CC-SA | CC-VNS | CC-SAVNS | CC-gurobi | CC-SA | CC-VNS |
| 15--3 | 600.00 | 4.48 | 3.17 | 3.27 | **1364.77** | 1442.51 | 1532.48 | 1388.95 | -1.77% | 3.71% | 9.37% |
| 18--3 | 600.00 | 3.62 | 3.52 | 3.87 | **1646.11** | 1684.44 | 1845.66 | 1656.64 | -0.64% | 1.65% | 10.24% |
| 24--3 | 3600.00 | 5.60 | 3.66 | 7.47 | **2300.41** | 2556.64 | 2627.84 | 2327.95 | -1.20% | 8.94% | 11.41% |
| 30--3 | 3600.00 | 15.68 | 4.74 | 12.34 | **3369.72** | 3521.73 | 3518.62 | 3386.27 | -0.49% | 3.85% | 3.76% |
| 40--4 | 3600.00 | 135.28 | 17.17 | 42.45 | ---- | 4968.40 | 5044.40 | 4636.50 | ---- | 6.68% | 8.09% |
| 48--4 | 3600.00 | 67.44 | 17.03 | 40.45 | ---- | 5537.63 | 5262.19 | 5165.07 | ---- | 6.73% | 1.85% |
| 60--4 | 3600.00 | 93.89 | 31.10 | 69.27 | ---- | 6003.78 | 6380.73 | 5753.30 | ---- | 4.17% | 9.83% |
| 80--4 | 3600.00 | 136.35 | 42.57 | 97.39 | ---- | 10545.58 | 10004.04 | 9086.75 | ---- | 13.83% | 9.17% |
| 90--3 | 3600.00 | 228.59 | 59.63 | 110.88 | ---- | 10574.87 | 10326.86 | 9896.28 | ---- | 6.42% | 4.17% |
| 120--3 | 3600.00 | 163.73 | 84.98 | 137.81 | ---- | 20033.85 | 16840.54 | 16039.62 | ---- | 19.94% | 4.76% |
| 150--3 | 6000.00 | 214.32 | 68.42 | 241.77 | ---- | 27702.04 | 25964.18 | 24929.93 | ---- | 10.01% | 3.98% |
| 180--3 | 6000.00 | 253.62 | 105.28 | 315.65 | ---- | 29655.85 | 28154.81 | 27451.94 | ---- | 7.43% | 2.50% |
| 120--4 | 6000.00 | 689.13 | 144.36 | 165.55 | ---- | 16043.49 | 17713.06 | 15242.88 | ---- | 4.99% | 13.95% |
| 160--4 | 6000.00 | 161.53 | 169.47 | 182.07 | ---- | 25799.91 | 25922.98 | 24934.14 | ---- | 3.36% | 3.81% |
| 200--4 | 6000.00 | 328.56 | 182.72 | 347.26 | ---- | 38791.81 | 37660.75 | 32073.59 | ---- | 17.32% | 14.84% |
| 240--4 | 6000.00 | 447.16 | 206.69 | 481.29 | ---- | 41349.27 | 43790.36 | 38742.82 | ---- | 6.30% | 11.53% |
| Average | 4125.00 | 187.51 | 71.53 | 141.17 | ---- | 15388.24 | 15161.84 | 13919.54 | ---- | 7.83% | 7.70% |
| Instances | Runtime(s) | | | | Obj | | | | Rel. Imp. | | |
| (n--m) | BOC-gurobi | BOC-SA | BOC-VNS | BOC-SAVNS | BOC-gurobi | BOC-SA | BOC-VNS | BOC-SAVNS | BOC-gurobi | BOC-SA | BOC-VNS |
| 15--3 | 600.00 | 4.66 | 6.48 | 4.80 | **1233.89** | 1406.19 | 1362.62 | 1253.29 | -1.57% | 10.87% | 8.02% |
| 18--3 | 600.00 | 8.56 | 7.72 | 3.66 | **1745.20** | 1817.27 | 1806.76 | 1767.16 | -1.26% | 2.76% | 2.19% |
| 24--3 | 3600.00 | 5.91 | 9.99 | 6.16 | **2450.87** | 2609.76 | 2527.20 | 2488.39 | -1.53% | 4.65% | 1.54% |
| 30--3 | 3600.00 | 21.96 | 13.25 | 27.69 | **2965.75** | 3187.40 | 3066.62 | 2968.77 | -0.10% | 6.86% | 3.19% |
| 40--4 | 3600.00 | 41.39 | 28.98 | 44.42 | ---- | 4323.84 | 4323.98 | 4101.41 | ---- | 5.14% | 5.15% |
| 48--4 | 3600.00 | 32.91 | 27.98 | 41.36 | ---- | 5395.17 | 4981.02 | 4648.00 | ---- | 13.85% | 6.69% |
| 60--4 | 3600.00 | 106.37 | 30.65 | 77.76 | ---- | 6273.35 | 6245.54 | 5817.46 | ---- | 7.27% | 6.85% |
| 80--4 | 3600.00 | 178.81 | 44.32 | 84.54 | ---- | 8122.09 | 7964.54 | 7537.50 | ---- | 7.20% | 5.36% |
| 90--3 | 3600.00 | 375.93 | 50.69 | 93.75 | ---- | 8200.64 | 8500.29 | 8016.37 | ---- | 2.25% | 5.69% |
| 120--3 | 3600.00 | 392.91 | 72.51 | 163.90 | ---- | 14138.61 | 14321.12 | 13409.76 | ---- | 5.16% | 6.36% |
| 150--3 | 6000.00 | 407.65 | 63.35 | 189.67 | ---- | 19110.16 | 19297.16 | 18597.80 | ---- | 2.68% | 3.62% |
| 180--3 | 6000.00 | 266.20 | 126.01 | 278.07 | ---- | 25603.26 | 26910.67 | 24206.55 | ---- | 5.46% | 10.05% |
| 120--4 | 6000.00 | 451.75 | 155.95 | 384.49 | ---- | 13877.20 | 17328.46 | 13559.46 | ---- | 2.29% | 21.75% |
| 160--4 | 6000.00 | 563.33 | 189.73 | 491.00 | ---- | 23828.40 | 26192.30 | 20616.91 | ---- | 13.48% | 21.29% |
| 200--4 | 6000.00 | 681.81 | 229.37 | 587.11 | ---- | 27786.00 | 31457.36 | 26481.69 | ---- | 4.69% | 15.82% |
| 240--4 | 6000.00 | 512.82 | 256.69 | 585.15 | ---- | 31723.06 | 37405.69 | 30356.11 | ---- | 4.31% | 18.85% |
| Average | 4125.00 | 253.31 | 82.10 | 191.47 | ---- | 12337.65 | 13355.71 | 11614.16 | ---- | 6.18% | 8.90% |
| Instances | Runtime(s) | | | | Obj | | | | Rel. Imp. | | |
| (n--m) | RBOC-gurobi | RBOC-SA | RBOC-VNS | RBOC-SAVNS | RBOC-gurobi | RBOC-SA | RBOC-VNS | RBOC-SAVNS | RBOC-gurobi | RBOC-SA | RBOC-VNS |
| 15--3 | 600.00 | 7.19 | 6.35 | 9.63 | **1333.89** | 1379.95 | 1481.03 | 1348.00 | -1.06% | 2.32% | 8.98% |
| 18--3 | 600.00 | 6.15 | 14.03 | 18.83 | **1711.43** | 1861.81 | 1783.38 | 1725.86 | -0.84% | 7.30% | 3.22% |
| 24--3 | 3600.00 | 8.69 | 27.67 | 37.90 | **2450.87** | 2640.67 | 2520.97 | 2467.63 | -0.68% | 6.55% | 2.12% |
| 30--3 | 3600.00 | 22.30 | 32.03 | 75.98 | **2949.96** | 3179.06 | 3085.86 | 2961.49 | -0.39% | 6.84% | 4.03% |
| 40--4 | 3600.00 | 68.24 | 29.11 | 83.80 | ---- | 4207.17 | 4257.07 | 3932.13 | ---- | 6.54% | 7.63% |
| 48--4 | 3600.00 | 55.72 | 32.44 | 64.28 | ---- | 5240.84 | 5470.92 | 4660.10 | ---- | 11.08% | 14.82% |
| 60--4 | 3600.00 | 119.76 | 40.12 | 145.84 | ---- | 6800.32 | 6143.57 | 6064.26 | ---- | 10.82% | 1.29% |
| 80--4 | 3600.00 | 153.16 | 38.73 | 138.81 | ---- | 7836.96 | 8164.62 | 6800.32 | ---- | 13.23% | 16.71% |
| 90--3 | 3600.00 | 200.49 | 40.98 | 194.02 | ---- | 8657.53 | 8928.64 | 8596.32 | ---- | 0.71% | 3.72% |
| 120--3 | 3600.00 | 389.91 | 88.80 | 367.45 | ---- | 16822.34 | 16226.28 | 13960.89 | ---- | 17.01% | 13.96% |
| 150--3 | 6000.00 | 326.26 | 120.44 | 312.13 | ---- | 20330.95 | 20843.57 | 17209.37 | ---- | 15.35% | 17.44% |
| 180--3 | 6000.00 | 329.11 | 112.84 | 383.11 | ---- | 29157.35 | 26886.47 | 26110.15 | ---- | 10.45% | 2.89% |
| 120--4 | 6000.00 | 226.45 | 115.86 | 278.87 | ---- | 18626.54 | 17324.16 | 15832.78 | ---- | 15.00% | 8.61% |
| 160--4 | 6000.00 | 314.92 | 119.78 | 381.46 | ---- | 24929.44 | 27110.78 | 24371.81 | ---- | 2.24% | 10.10% |
| 200--4 | 6000.00 | 261.09 | 224.69 | 351.47 | ---- | 30364.87 | 31321.70 | 28594.33 | ---- | 5.83% | 8.71% |
| 240--4 | 6000.00 | 308.65 | 218.17 | 399.27 | ---- | 33368.95 | 39415.51 | 31712.38 | ---- | 4.96% | 19.54% |
| Average | 4125.00 | 174.88 | 78.88 | 202.68 | ---- | 13462.80 | 13810.28 | 12271.74 | ---- | 8.51% | 8.99% |

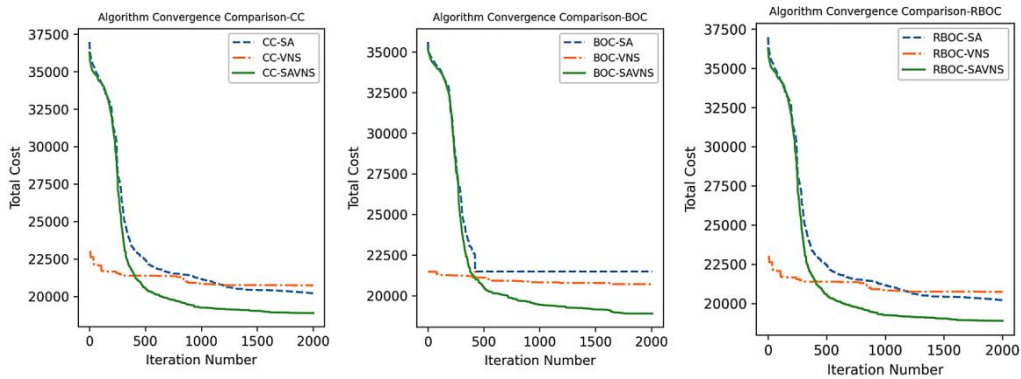

Figure 6 Caption: Convergence results of heuristic approaches for collaborative models

Figure 6 Alt Text: These figures illustrate the convergence of hybrid SAVNS-based heuristics for three collaborative strategies, in comparison to two other heuristics: CC-SAVNS convergence ((left figure), BOC-SAVNS convergence (middle figure), and RBOC-SAVNS convergence (right figure).

### *5.2 The efficiency of the collaborative strategies*

*5.2.1 Experimental results*

We evaluate three collaborative strategies across 4 MDVRPTW instances labeled as "*n-m-s*", indicating customer, depot, and strategy variations. Euclidean distance is used to calculate the distance between nodes, and demands as well as service times are randomly generated within specified ranges as in Section 5.1. For the RBOC strategy, we assume a discount $\alpha$ of 0.8 for travel unit cost after distribution and a highway between distribution centers if their distance is less than 60 km. Each experiment is repeated 5 times, and the average outcomes are reported. The stability of our hybrid heuristics is verified by calculating the ratio between standard deviations and averages (*Std/Avg*) and relative differences between averages (*Mre*) for each experiment, as shown in Table 3.

Table 3 displays cost items with their minimum values (bolded) for each instance. It shows that BOC and RBOC require fewer vehicles than CC, thereby reducing fix costs. However, the traditional belief that OVRP achieves lower transportation costs than $C_{LU}VRP$ is not witnessed when vehicle flow equilibrium at each depot is achieved. Although CC incurs lower transportation cost, it is offset by other cost items. In most cases, BOC and RBOC generate less total cost than CC. Furthermore, BOC and RBOC consistently yield lower penalty costs, indicating higher customer satisfaction. CC exhibits a slight advantage over the alternative strategies in terms of good loss and cooling costs.

Table 3. Performance of different collaborative strategies on various cost items

| Instances n-m-s | Cost | | | | | | | | Mre | Std/Avg |
|---|---|---|---|---|---|---|---|---|---|---|
| | Total | Fix | Transportation | Penalty | Good Loss | CO2-emission | Cooling | Re-balancing (From, To, #Veh) | | |
| 72-4-CC | 33092.31 | 5000 | **16409.68** | 1832.59 | 691.58 | 7411.29 | 1747.17 | -- | 1.15% | 0.010 |
| 72-4-BOC | **31804.77** | 4500 | 17025.27 | **1397.02** | 696.56 | **6402.59** | 1783.32 | -- | 2.92% | 0.033 |
| 72-4-RBOC | 32190.48 | 4500 | 16917.14 | 1439.29 | **691.54** | 6899.79 | **1742.73** | 0.00 -- | 3.82% | 0.029 |
| 72-6-CC | 31675.63 | 6000 | **15750.04** | 2491.01 | **678.43** | 5062.44 | 1693.71 | -- | 0.25% | 0.003 |
| 72-6-BOC | **31774.12** | 4500 | 16784.41 | **1392.39** | 693.37 | **6650.26** | 1753.69 | -- | 3.07% | 0.018 |
| 72-6-RBOC | 32687.24 | 4500 | 16818.42 | 1455.18 | 688.92 | 7151.59 | **1732.64** | 340.49 (D5, D3, 1Veh) | 0.11% | 0.013 |
| 144-4-CC | 62637.58 | 8500 | **27286.50** | 3102.41 | 1450.74 | 18825.97 | 3471.97 | -- | 1.49% | 0.011 |
| 144-4-BOC | 61009.27 | 7500 | 29250.50 | **2308.80** | 1461.15 | 16953.06 | 3535.75 | -- | 0.76% | 0.006 |
| 144-4-RBOC | **60060.70** | 7500 | 28086.76 | 2331.16 | **1433.18** | **16550.12** | **3447.06** | 712.43 (D2, D3, 1Veh) (D4, D3, 1Veh) | 3.53% | 0.030 |
| 144-6-CC | 62112.93 | 8000 | **26138.99** | 2640.04 | 1447.34 | 20420.92 | 3465.64 | -- | 0.51% | 0.005 |
| 144-6-BOC | **58075.36** | 7500 | 26532.43 | 2395.10 | **1443.64** | 16748.63 | **3455.57** | -- | 3.03% | 0.017 |
| 144-6-RBOC | 58854.09 | 7500 | 27746.57 | **2269.96** | 1463.55 | **15934.10** | 3498.41 | 441.50 (D2, D5, 1Veh) | 1.55% | 0.019 |

An intriguing finding is that RBOC sometimes can achieve vehicle flow equilibrium at each depot during distribution without the need for re-balancing operations. Meanwhile, in cases where re-balancing is performed, only a few vehicle transfers between depots are required, incurring a marginal cost that enhances the feasibility of RBOC. Although this may be attributed to a large discount value $\alpha$ we used, indicating a significant reduction in travel unit cost after distribution, our sensitivity analysis confirms that changes in this coefficient do not affect our conclusion.

*5.2.2 Sensitivity analysis*

In this section, we investigate the impact of specific parameters on the larger-scale instance (144 customer-6 depots) to verify our previous findings. We examine the sensitivity of the carbon emission coefficient ($\lambda$), the refrigeration cost coefficient ($c_1$), and the discount of travel unit cost ($\alpha$). In particular, we evaluate the effect of the discount $\alpha$ on the re-balancing operation costs in the RBOC strategy.

The above panels in Figure 7 demonstrate that increasing the value of $\lambda$ results in higher CO2-emission costs. For small values of $\lambda$ [0.25, 0.5, 1], all strategies have similar CO2-emission costs. However, when $\lambda$ is greater than 1.5, CC incurs higher costs compared to BOC and RBOC, and the gap between them widens. Changes in $\lambda$ do not significantly affect penalty costs, as shown in the top right panel of Figure 7. It is important to emphasize that the selection of vehicle fuel exclusively affects the carbon emissions associated with the collaborative strategy, without altering the earlier finding that BOC and RBOC outperform CC in environmental impacts. The bottom two panels of Figure 7 reveal that CC consistently results in the highest CO2-emission cost and penalty cost, regardless of changes in $c_1$, while the penalty costs of BOC and RBOC remain consistent.

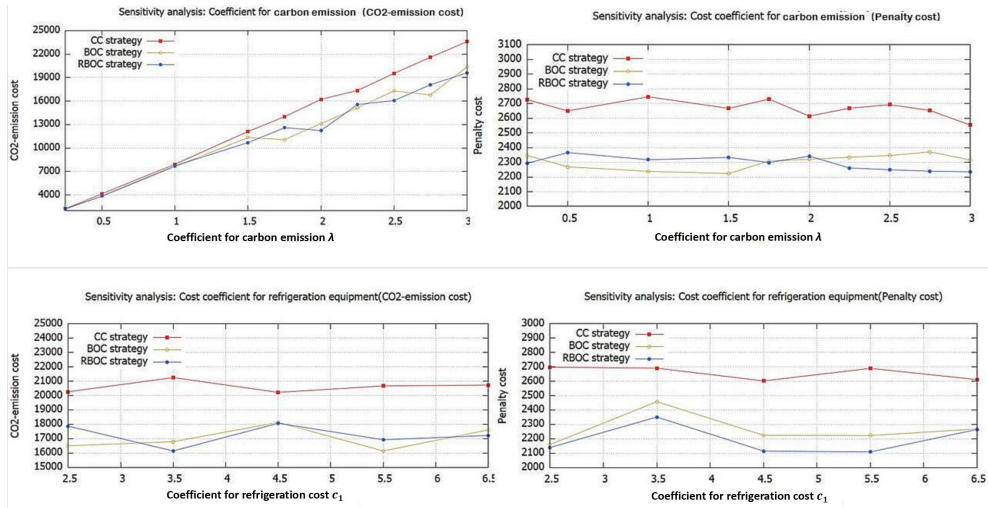

Figure 7 Caption: Sensitivity analysis of $\lambda$ and $c_1$ coefficients for CO2-emission & Penalty costs

Figure 7 Alt Text: The upper panels show the performance of CO2-emission cost (left) and Penalty cost (right) with the variety of $\lambda$; The lower panels show the performance of CO2-emission cost (left) and Penalty cost (right) with the variety of $c_1$.

Table 4. Improvements of various cost items in different $\alpha$ (RBOC vs CC | RBOC vs BOC)

| Cost items | Strategy | Value | Discount $\alpha$ | | |
|---|---|---|---|---|---|
| | | | Low | Med | High |
| Total | RBOC (Mean) | - | 61772.95 | 61016.85 | 60897.38 |
| | vs CC | 62112.93 | 0.55% | 1.76% | 1.96% |
| | vs BOC | 58075.36 | -6.37% | -5.06% | -4.86% |
| Transportation | RBOC (Mean) | - | 28654.73 | 28676.85 | 28071.31 |
| | vs CC | 26138.99 | -9.62% | -9.71% | -7.39% |
| | vs BOC | 26532.43 | -8.00% | -8.08% | -5.80% |
| Penalty | RBOC (Mean) | - | 2191.15 | 2204.46 | 2277.55 |
| | vs CC | 2640.04 | 17.00% | 16.50% | 13.73% |
| | vs BOC | 2395.10 | 8.52% | 7.96% | 4.91% |
| Good Loss | RBOC (Mean) | - | 1448.68 | 1450.62 | 1451.18 |
| | vs CC | 1447.34 | 0.09% | 0.23% | 0.27% |
| | vs BOC | 1443.64 | 0.35% | 0.48% | 0.52% |
| CO2 Emission | RBOC (Mean) | - | 17569.44 | 17120.13 | 17185.39 |
| | vs CC | 20420.92 | 13.96% | 16.16% | 15.84% |
| | vs BOC | 16748.63 | -4.90% | -2.22% | -2.61% |
| Cooling | RBOC (Mean) | - | 3526.93 | 3481.04 | 3491.27 |
| | vs CC | 3465.64 | -1.77% | -0.44% | -0.74% |
| | vs BOC | 3455.57 | -2.07% | -0.74% | -1.03% |
| Re-balancing | RBOC (Mean) | - | 923.56 | 586.85 | 852.34 |

Finally, we examine how the discount $\alpha$ affects the performance of RBOC. We divide the discount value into low [0,0.35], med [0.4,0.65] & high [0.7,1.0] levels and conduct five replications. The average results, displayed in Table 4, provide a comparative analysis of cost items and savings enhancements in comparison to CC and BOC. It shows that the change of $\alpha$ has a certain impact on RBOC manifestation, i.e., when the government subsidy is stronger (larger value of $\alpha$), RBOC is able to generate lower total costs and tends to perform better in partial cost items, such as transportation cost, CO2-emission cost and cooling cost. This indicates that government subsidies are valid and provide financial benefits to certain extent for RBOC. However, from the perspective of strategy comparison, we find that variations in $\alpha$ do not change the comparative results of RBOC, which also indicates that RBOC is practical and effective, with little effect on its sustainable performance as $\alpha$ varies. This suggests that RBOC allows re-balancing of vehicle flow among depots after distribution without relying on strong policy subsidies. It is noteworthy that irrespective of the $\alpha$ value, RBOC incurs a minimal re-balancing cost, amounting to less than 2% of the total distribution cost, affirming its practical viability.

*5.3 Case study*

*5.3.1 The benefits of collaboration*

In this section, we conduct a case study in Shanghai, China involving 114 medical centers and 6 distribution centers to demonstrate the sustainable benefits of our collaborative strategies. We use Google map API to obtain location data, see Figure 8, and assume that distribution centers have various transport resources (a random positive integer between 6 and 10). Taking vaccine distribution during the pandemic as a case, the vaccine demand for each medical center ranges from 1 to 25 boxes (1 box contains 500 doses of vaccine). Diesel vehicles are used with a carbon emission coefficient $\lambda$ of 2.61kgCO2/L, and a maximum load capacity of 80 boxes. We set the value of $\alpha$ at a medium level (0.4) and assume that distribution centers are within 45km of each other existing a highway. Other parameters are set similarly as Section 5.2. The daily vaccine demand of a single medical center $I_n \in N$ is set as $\Omega_n$. In the stand-alone scenario, each medical center orders $\omega_1$ and $\omega_2$ boxes of vaccines separately from the two nearest distribution centers ($\omega_1 + \omega_2 = \Omega_n$). Four scenarios are analyzed: stand-alone, CC, BOC, and RBOC strategies. Each scenario is replicated five times and the average results are reported.

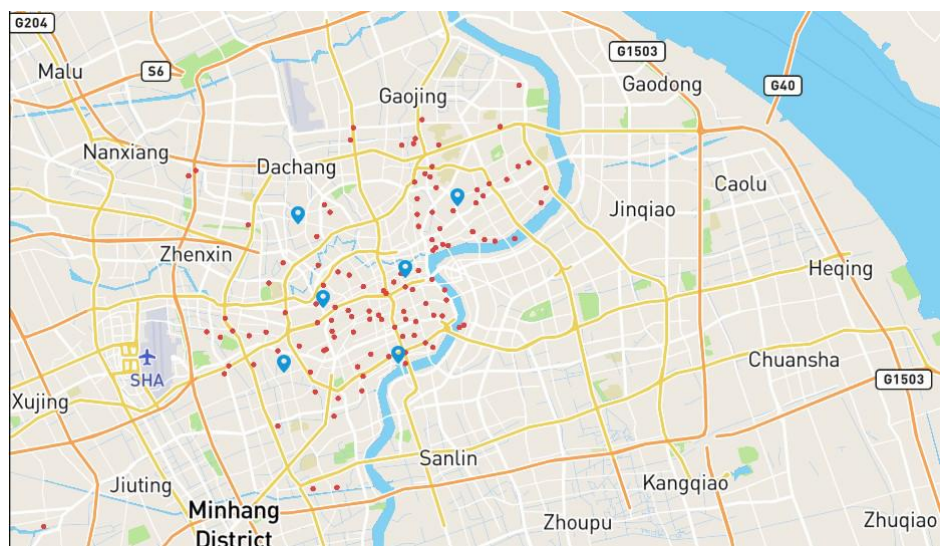

Figure 8 Caption: Medical centers (red set points) and distribution centers (blue set points) in Shanghai China

Figure 8 Alt Text: 114 medical centers (red set points) and 6 distribution centers (blue set points) in Shanghai China.

Besides the comparison of cost items, we evaluate the performance in terms with loading rate (*LR*) and customer satisfaction (*CS*). Loading rate refers to the efficiency of vehicles in carrying goods, and a full loading rate (*FLR*) indicates the extent to which vehicles are fully utilized. Customer satisfaction is measured by delivery punctuality and is calculated based on a fuzzy appointment time function (Du and Li 2020). If vaccines arrive moderately before the earliest time window (set as $et^*$), CS is set to 1, and it decreases linearly between the earliest and latest time windows ($et \& lt$). Otherwise, customer satisfaction is 0. We calculate customer satisfaction $CS_i$ of customer $i$ visited by vehicle $h \in H$ by equation (24) with the arrival time $A_{ih}$.

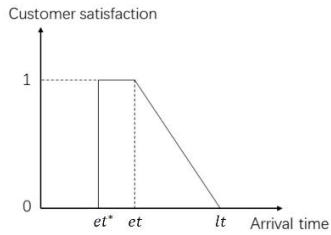

$$CS_i = \begin{cases} 1, & et^* \leq A_{ih} \leq et \\ 1 - \frac{A_{ih} - et}{lt - et}, & et \leq A_{ih} \leq lt \\ 0, & \text{otherwise} \end{cases} \quad (24)$$

We also measure the percentage of vehicles arriving too early or too late, called as the early arrival ratio (*EAR*) and tardiness ratio (*TR*). Considering the perishability of cold-chain drugs, a higher *EAR* is better for vaccine distribution, while a smaller *TR* is preferred.

Table 5 presents the results of various indicators. The upper part compares the outcomes of the four scenarios, demonstrating significant cost savings in all collaborative strategies, with over 35% savings in total costs on average. The middle part shows the average improvements in cost items relative to the stand-alone scenario, revealing substantial savings in transport resources, transportation, CO2 emissions, and cooling costs, with average savings exceeding 40% for each cost item. Friedman's test (Demsar 2006) confirms significant performance differences among the three strategies ($p<0.05$). BOC and RBOC perform comparably in total and fix costs, with BOC having lower penalty costs and RBOC achieving superior CO2 emissions reduction. On average, BOC and RBOC yield 8% higher cost savings than CC for each of the aforementioned cost items. Although CC outperforms the other strategies in transportation, good loss, and cooling cost, its overall performance is not superior due to subtle differences in these costs among the three strategies. Details are available in Figure 9.

Regarding other indicators compared with stand-alone (see bottom part of Table 5), all collaborative strategies achieve higher *LR*, exceeding 88%, with BOC and RBOC reaching 98.13%. RBOC achieves a 50% *FLR*, surpassing BOC's 35%. Customer satisfaction improves for all strategies, with BOC bringing the greatest increase at 8.06%, followed by RBOC at 4.26%, and CC at 3.76% in *CS*. The collaborative strategies also achieve higher *EAR*, with CC achieving 35.98%, BOC achieving 31.17%, and RBOC achieving 33.77%. Additionally, the *TR* is extremely low in either scenario, with CC and BOC avoiding late deliveries entirely.

Table 5. Computational results in different scenarios (Stand-alone vs Collaboration)

| Scenarios | Cost | | | | | | |
|---|---|---|---|---|---|---|---|
| | Total | Fix | Transportation | Penalty | Good Loss | CO2 Emission | Cooling |
| Stand-alone | 40725.88 | 12000.00 | 10924.43 | 3416.53 | 992.82 | 9136.24 | 4255.87 |
| CC | 26508.29 | 11000.00 | 6094.87 | 3432.34 | 948.28 | 3170.32 | 1862.48 |
| BOC | 25015.16 | 10000.00 | 6632.53 | 3046.02 | 999.95 | 2350.33 | 1986.34 |
| RBOC | 25096.94 | 10000.00 | 6783.07 | 3177.46 | 992.02 | 2175.79 | 1968.60 |
| | Collaboration vs Stand-alone: Cost savings (Decrease) | | | | | | |
| Scenarios | Total | Fix | Transportation | Penalty | Good Loss | CO2 Emission | Cooling |
| CC | 34.91% | 8.33% | 44.21% | -0.46% | 4.49% | 65.30% | 56.24% |
| BOC | 38.58% | 16.67% | 39.29% | 10.84% | -0.72% | 74.27% | 53.33% |
| RBOC | 38.38% | 16.67% | 37.91% | 7.00% | 0.08% | 76.19% | 53.74% |
| Mean | 37.29% | 13.89% | 40.47% | 5.79% | 1.28% | 71.92% | 54.44% |
| | | | | | | $p = 0.022$ * | * $p<0.05$ ** $p<0.01$ |

| Scenarios | LR | FLR | Increase in LR (vs Stand-alone) | CS | EAR | TR | Increase in CS |
|---|---|---|---|---|---|---|---|
| Stand-alone | 78.33% | 0.00% | | 61.15% | 25.40% | 0.23% | |
| CC | 88.78% | 31.82% | 13.35% | 63.45% | 35.98% | 0.00% | 3.76% |
| BOC | 98.13% | 35.00% | 25.28% | 66.08% | 31.17% | 0.00% | 8.06% |
| RBOC | 98.13% | 50.00% | 25.28% | 63.76% | 33.77% | 0.65% | 4.26% |

*LR*: Loading Rate, *FLR*: Full Loading Ratio, *CS*: Customer Satisfaction, *EAR*: Early Arrival Ratio, *TR*: Tardiness Ratio

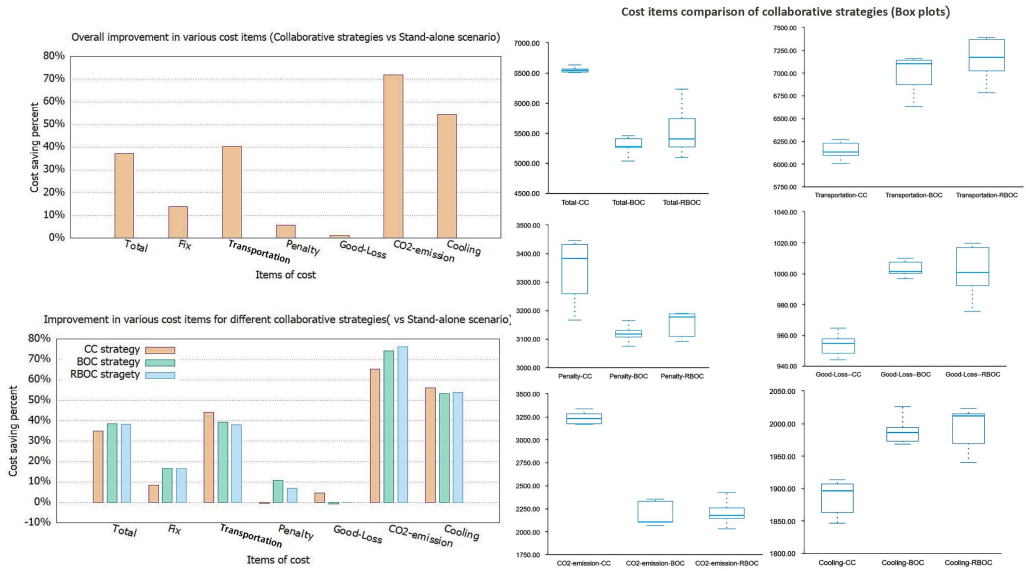

Figure 9 Caption: (Left) Cost improvement comparison; (Right) Box plots of cost items for three Collaborative strategies

Figure 9 Alt Text (Long Description): The left graphs illustrate cost savings benefits of collaboration vs stand-alone in each cost item (top) and cost savings advantages of different collaborative strategies vs stand-alone in each cost item (bottom). The right graphs depict box plots of three collaborative strategies for different cost items across five experiments.

*5.3.2 The impact of departure time*

As detailed in Section 3.2, our initial experiments utilized a uniform departure time of 9:00 AM for all vehicles from the depots. In this section, we investigate the impact of different departure time policies on the performance of the three collaborative strategies. Leveraging data from the case study, we examine and compare various departure time configurations. Notably, Figure 2 indicates the occurrence of early congestion period (lower travel speed) between 9:00 AM and 10:00 AM. To explore the potential enhancement of logistics efficiency across the collaborative strategies, we propose both fixed time policies and flexible time policies, aiming to determine whether escaping from the early congestion period can yield beneficial outcomes:

- Fixed time policies: all vehicles depart from the depot at a specific predetermined time point, selected from the set [9:00, 9:10, 9:20, 9:30, 9:40, 9:50, 10:00].

- Flexible time policies: the departure time of each vehicle from the depot follows a uniform distribution within the intervals [9:00, 9:30], [9:30, 10:00], or [9:00, 10:00].

For each departure time configuration in these two policies, we conduct five iterations and the average outcomes are reported. A total of 150 experiments have been conducted, and the detailed results are presented in Table A2 (see Appendix 2). According to the outcomes, regardless of the departure time policies, the three strategies consistently demonstrate stable performances in terms of transport resources utilization (fix cost), transportation distance (transportation cost), and CO2-emission cost. This indicates that the main findings obtained from the comparison of collaborative strategies remain valid.

Furthermore, it is worth noting that different departure time policies do have an impact on the performance of each strategy. Figure 10 illustrates that under fixed time policies (depicted in the left panel for each cost item), a later set-off time leads to lower costs in most of the considered categories, i.e., total cost, penalty cost, good loss cost, and cooling cost, for each strategy. Nevertheless, the influence of departure time does not change our previous conclusions that BOC and RBOC exhibit lower total cost and CO2 emissions compared to CC, while CC demonstrates a slight advantage in other cost items. These conclusions are also evident when considering the flexible time policies, as shown in the right panel of each cost item in Figure 10. We can observe that granting vehicles the flexibility to select a later departure time (after 9:30 AM) can significantly improve logistics efficiency and lead to cost savings in the aforementioned cost items.

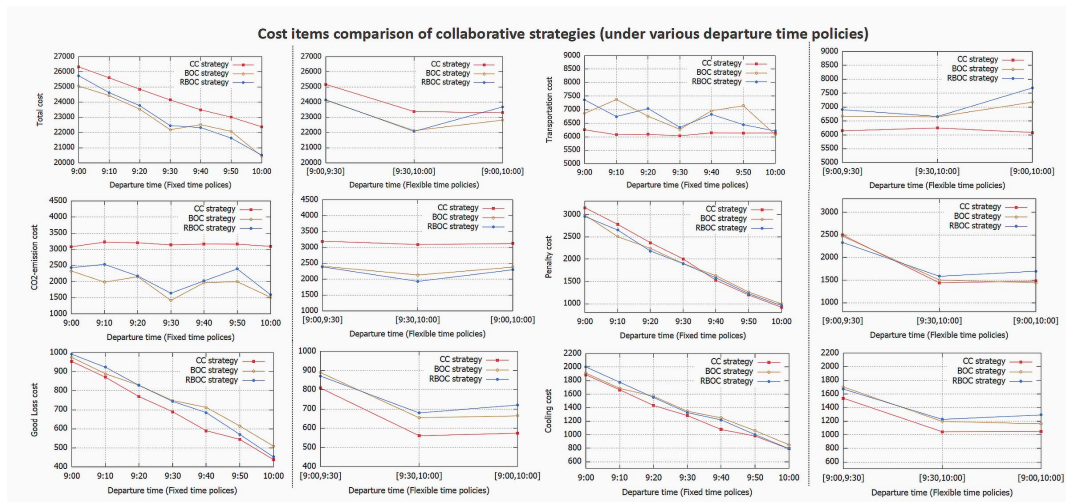

Figure 10 Caption: Cost items comparison of collaborative strategies (under various departure time policies).

Figure 10 Alt Text: Each cost item (i.e., total, transportation, penalty, good loss, CO2-emission and cooling cost) contains two figure panels (left panel shows the results of fixed time policies and right panel shows the results of flexible time policies).

### 6. Conclusion and managerial Insights

In recent years, the global refrigerated logistics market has seen surging demand for perishable goods, but also poses serious sustainability threats due to limited transport resources, excessive carbon emissions, and high operating costs. Taking the instance of pharmaceutical refrigeration logistics, this research presents three sustainable collaborative strategies aimed at achieving vehicle flow equilibrium at each depot, ensuring long-term operational stability in logistics performance. The problem is regarded as a multi-depot vehicle routing problem with time windows (MDVRPTW) and the proposed sustainable distribution strategies are based on the $C_{LU}$VRP and improved OVRP. Our strategies focus on ensuring vehicle flow equilibrium at each depot during the collaboration through diverse approaches.

To this end, specific hybrid heuristics based on Simulated Annealing and Variable Neighborhood Search (SAVNS) algorithms are proposed and we verify their effectiveness through algorithmic experiments. The comparative studies yield several intriguing findings and demonstrate the practical applicability of our models. The case study on vaccine distribution further confirms the advantages of the proposed collaborative strategies over a stand-alone scenario in terms of TBL performance. This study provides the following managerial insights for decision makers, highlighting the benefits of implementing sustainable collaborative strategies in refrigerated logistics:

- Our study demonstrates that horizontal collaboration among distribution centers can lead to sustainable benefits in terms of TBL. However, different collaborative strategies exhibit various logistics performances, making them suitable for different distribution scenarios:
  - When overall transport resources are limited, the BOC or RBOC strategies are recommended. These strategies utilize fewer vehicles for delivery and achieve better transport resource utilization. On the other hand, the CC strategy is ideal for reducing the total distance of distribution tasks, enhancing employee-friendliness, and convenience during implementation.
  - For decision-makers prioritizing environmental benefits, the BOC strategy is suggested. It offers efficient deliveries and mobility while ensuring vehicle flow equilibrium at each depot during distribution. The RBOC strategy is also a viable option for those addressing distribution and sustainability concerns separately, as it demonstrates comparable sustainable performance and cost-efficiency even without strong government subsidies.
- Our extended analysis in this case study underscores the significance of avoiding early congestion periods to improve the logistics efficiency of the proposed strategies within real distribution scenarios. This consideration becomes particularly crucial when factoring in time-dependent travel speeds, which exert a substantial influence on refrigerated logistics. The findings show that strategically postponing the departure time from the depot during the early congestion period can effectively lower the total distribution cost, enhance consumer satisfaction, and save costs in terms of good loss and cooling expenses. Importantly, the essential features of the three strategies remain unaffected by variations in the departure time.

Our study motivates future research to incorporate more sustainable concerns in the refrigerated logistics research. Notably, the current study has some limitations. First, it only considers the pharmaceutical refrigerated distribution with a homogenous fleet of vehicles. Due to the diversity of cold-chain products, this study can be extended to other industries such as meat, agri-food, etc., while considering heterogeneous refrigerated trucks. Second, all of our computational experiments use deterministic demands. To better match the complexity and variability of realistic distribution scenarios, taking stochastic demands into account in our study will be an interesting topic to explore.

**Data availability statement**

The data that support the findings of this study are available from the corresponding author upon reasonable request.

**Disclosure statement**

No potential conflict of interest was reported by the authors.

**Funding**

This research was supported by the National Natural Science Foundation of China (No. 71971156 and No72371188), the Fundamental Research Funds for the Central Universities (No. 22120210241).

## Appendices

### Appendix 1 – Linearization of the basic model

We introduce some sets of auxiliary variables to linearize the proposed model. The penalty cost function $\sum_{\pi \in \Pi} \sum_{i \in N} \sum_{h \in H} Y_{ih}^{\pi}(c_2(et_i - A_{ih})^+ + c_3(A_{ih} - lt_i)^+)$ has a nonlinear form '$(\blacksquare)^+$' and a multiplication of decision variables ($Y_{ih}^{\pi}$ with $A_{ih}$).

At first, to address the nonlinear form $c_2(et_i - A_{ih})^+$, we define non-negative variables $\vartheta_{eih}^+$ and $\vartheta_{eih}^-$, and add a constraint $et_i - A_{ih} = \vartheta_{eih}^+ - \vartheta_{eih}^-, \forall i \in N, \forall h \in H$. Then, we replace this nonlinear part in the penalty cost function with $\vartheta_{eih}^+$ and make it become linear, i.e., $c_2 \vartheta_{eih}^+$. Here $\vartheta_{eih}^+$ and $\vartheta_{eih}^-$ can be defined as two non-negative continuous variables, and as the objective function suggests, $\vartheta_{eih}^+$ should be as less as possible. Therefore, we denote that if the value of the function $et_i - A_{ih}$ is negative, $\vartheta_{eih}^+ = 0$, $\vartheta_{eih}^+ = et_i - A_{ih}$ otherwise. Similarly, we linearize the nonlinear form $c_3(A_{ih} - lt_i)^+$ by introducing another two non-negative variables $\vartheta_{lih}^+$ and $\vartheta_{lih}^-$. Then a constraint $A_{ih} - lt_i = \vartheta_{lih}^+ - \vartheta_{lih}^-, \forall i \in N, \forall h \in H$ is added. We replace $(A_{ih} - lt_i)^+$ with $\vartheta_{lih}^+$ and thus this nonlinear part in the penalty cost function becomes linear, i.e., $c_3 \vartheta_{lih}^+$. Likewise, $\vartheta_{lih}^+$ and $\vartheta_{lih}^-$ can be defined as two non-negative continuous variables and $\vartheta_{lih}^+$ should be as less as possible. If the value of the function $A_{ih} - lt_i$ is negative, $\vartheta_{lih}^+ = 0$, $\vartheta_{lih}^+ = A_{ih} - lt_i$ otherwise.

With the linearization of the '$(\blacksquare)^+$' part in the penalty cost function, the original function can be transformed into $\sum_{\pi \in \Pi} \sum_{i \in N} \sum_{h \in H} Y_{ih}^{\pi}(c_2 \vartheta_{eih}^+ + c_3 \vartheta_{lih}^+)$. Now we proceed to linearize the nonlinear form $Y_{ih}^{\pi} \vartheta_{eih}^+$ and $Y_{ih}^{\pi} \vartheta_{lih}^+$. We define two non-negative auxiliary variables: $\mu_{eih}^{\pi} = Y_{ih}^{\pi} \vartheta_{eih}^+$ and $\mu_{lih}^{\pi} = Y_{ih}^{\pi} \vartheta_{lih}^+$. Then the penalty cost function can be linearized as $\sum_{\pi \in \Pi} \sum_{i \in N} \sum_{h \in H} (c_2 \mu_{eih}^{\pi} + c_3 \mu_{lih}^{\pi})$. We denote that $\mu_{eih}^{\pi}$ (resp. $\mu_{lih}^{\pi}$) takes value $\vartheta_{eih}^+$ (resp. $\vartheta_{lih}^+$) if vehicle $h$ departs from node $i$ at time period $\pi$, and 0 otherwise. After the linearization, our basic model can be rewritten as follows:

$$min \sum_{i \in M, j \in N} \sum_{h \in H} f X_{ijh} + \sum_{i \in PN, j \in PN^*} \sum_{h \in H} c_0 d_{ij} X_{ijh} + \sum_{i \in PN, j \in PN^*} \sum_{h \in H} E_{jh}^{\omega_j} X_{ijh} +$$
$$\sum_{i \in PN, j \in N} \sum_{h \in H} (c_1 + \beta \omega_j)(t_{ijh} + TT_j) X_{ijh} + \sum_{\pi \in \Pi} \sum_{i \in N} \sum_{h \in H} (c_2 \mu_{eih}^{\pi} + c_3 \mu_{lih}^{\pi}) \tag{A1}$$

Subject to

$$(2)\text{-}(17);$$

$$et_i - A_{ih} = \vartheta_{eih}^+ - \vartheta_{eih}^- \quad \forall i \in N, \forall h \in H \tag{A2}$$

$$A_{ih} - lt_i = \vartheta_{lih}^+ - \vartheta_{lih}^- \quad \forall i \in N, \forall h \in H \tag{A3}$$

$$\mu_{eih}^{\pi} \leq \vartheta_{eih}^{+} \quad \forall i \in N, \forall h \in H, \forall \pi \in \Pi \quad (A4)$$

$$\mu_{eih}^{\pi} \leq WY_{ih}^{\pi} \quad \forall i \in N, \forall h \in H, \forall \pi \in \Pi \quad (A5)$$

$$\mu_{lih}^{\pi} \leq \vartheta_{lih}^{+} \quad \forall i \in N, \forall h \in H, \forall \pi \in \Pi \quad (A6)$$

$$\mu_{lih}^{\pi} \leq WY_{ih}^{\pi} \quad \forall i \in N, \forall h \in H, \forall \pi \in \Pi \quad (A7)$$

$$\vartheta_{eih}^{+}, \vartheta_{eih}^{-}, \vartheta_{lih}^{+}, \vartheta_{lih}^{-} \geq 0, \quad \forall i \in N, \forall h \in H \quad (A8)$$

$$\mu_{eih}^{\pi}, \mu_{lih}^{\pi} \geq 0, \forall i \in N, \forall h \in H, \forall \pi \in \Pi \quad (A9)$$

The objective function (A1) is a linear function. Constraint (A2) calculates the variable $\vartheta_{eih}^{+}$ and Constraint (A3) calculates the variable $\vartheta_{lih}^{+}$. Constraints (A4)-(A5) show the relationship between variables $\mu_{eih}^{\pi}$, $\vartheta_{eih}^{+}$ and $Y_{ih}^{\pi}$. Constraints (A6)-(A7) show the relationship between variables $\mu_{lih}^{\pi}$, $\vartheta_{lih}^{+}$ and $Y_{ih}^{\pi}$. Finally. Constraints (A8)-(A9) define the domain of decision variables.

## Appendix 2 – Supplementary tables

Table A1. Settings of parameters in various scenarios

| Parameters | Definition | Value/Range | Unit | Source |
|---|---|---|---|---|
| $f$ | Fixed cost of operating a vehicle | 500 | $ | Assumption |
| $\Pi$ | Set of time periods | {[0, 60), [60, 120), [120, 180), [180, 240), [240, 300), [300, 360), [360, 420), [420, 480]} | min | Assumption |
| $v_{\pi}$ | Average speed of a vehicle at time period $\pi$ | (10, 15, 15, 30, 30, 15, 15, 10) | km/h | Zhang et al. (2019) |
| $c_0$ | Travel cost coefficient | 10 | $/km | Assumption |
| $\alpha$ | Travel cost discount (after distribution) | [0,1] | Ratio | Assumption |
| $c_1$ | Cooling cost coefficient | 4.5 | $/km | Assumption |
| $(c_2, c_3)$ | Penalty cost coefficient for time window violations (overly early/late) | (5,10) | $/h | Assumption |
| $\lambda$ | Carbon emission coefficient | 2.61 | kgCO2/L | Dura Ja (2013) |
| $e$ | Carbon cost coefficient | 0.1 | $/kgCO2 | Bao et al. (2018) |
| $\beta$ | Good loss coefficient (distribution/unloading) | 0.005 | $/unit | Zhang et al. (2019) |
| **Hybrid SAVNS Settings** | | | | |
| $(\eta, \theta, \xi)$ | Weighting factors of insertion cost function in PFIH | (0.7, 0.2, 0.1) | Ratio | Zhang et al. (2019) |
| $T_0$ | Initial temperature in SA | 5000 | °C | Assumption |
| $\delta$ | Coefficient of temperature cooling in SA | 0.98 | Ratio | Zhang et al. (2019) |
| $T^*$ | Termination temperature in SA | 1 | °C | Assumption |
| $K_{max}$ | Number of variable neighborhoods in VNS | 8 | Integer | Zhang et al. (2019) |

Table A2. Performance of different departure time policies on various cost items

| Cost items | Strategy | Fixed Time | | | | | | | | Flexible Time | | | |
|---|---|---|---|---|---|---|---|---|---|---|---|---|---|
| | | 9:00 | 9:10 | 9:20 | 9:30 | 9:40 | 9:50 | 10:00 | Average | [9:00,9:30] | [9:30,10:00] | [9:00,10:00] | Average |
| Total | CC | 26330.13 | 25607.94 | 24861.99 | 24151.17 | 23503.32 | 23016.98 | 22385.87 | **24265.34** | 25184.99 | 23386.82 | 23312.55 | **23961.45** |
| | BOC | 25066.10 | 24439.50 | 23548.17 | 22192.18 | 22521.82 | 22084.16 | 20458.71 | **22901.52** | 24134.71 | 22138.59 | 22823.33 | **23032.21** |
| | RBOC | 25746.19 | 24632.15 | 23777.53 | 22463.62 | 22336.58 | 21641.73 | 20520.47 | **23016.90** | 24169.87 | 22081.59 | 23698.23 | **23316.56** |
| Fix | CC | 11000 | 11000 | 11000 | 11000 | 11000 | 11000 | 11000 | **11000** | 11000 | 11000 | 11000 | **11000** |
| | BOC | 10000 | 10000 | 10000 | 10500 | 10000 | 10000 | 10500 | **10143** | 10000 | 10000 | 10000 | **10000** |
| | RBOC | 10000 | 10000 | 10000 | 10500 | 10000 | 10000 | 10500 | **10143** | 10000 | 10000 | 10000 | **10000** |
| Transportation | CC | 6257.40 | 6072.52 | 6089.19 | 6034.98 | 6142.93 | 6132.82 | 6136.45 | **6123.76** | 6146.33 | 6244.96 | 6083.71 | **6158.33** |
| | BOC | 6863.45 | 7376.39 | 6756.66 | 6273.46 | 6959.86 | 7144.73 | 6079.83 | **6779.20** | 6665.50 | 6647.26 | 7176.45 | **6829.74** |
| | RBOC | 7361.94 | 6746.67 | 7041.51 | 6348.50 | 6822.61 | 6446.96 | 6212.99 | **6711.60** | 6904.24 | 6658.91 | 7688.70 | **7083.95** |
| Penalty | CC | 3153.08 | 2778.19 | 2369.12 | 2006.00 | 1528.63 | 1198.86 | 922.57 | **1993.78** | 2496.12 | 1440.26 | 1479.56 | **1805.31** |
| | BOC | 2988.20 | 2503.52 | 2241.33 | 1903.22 | 1629.23 | 1263.76 | 999.05 | **1932.61** | 2468.57 | 1505.04 | 1439.63 | **1804.41** |
| | RBOC | 2955.72 | 2652.87 | 2182.65 | 1900.63 | 1583.77 | 1226.96 | 960.26 | **1923.27** | 2333.09 | 1584.46 | 1697.52 | **1871.69** |
| Good Loss | CC | 953.08 | 870.45 | 768.92 | 689.69 | 590.35 | 544.59 | 439.03 | **693.73** | 808.75 | 561.16 | 573.89 | **647.94** |
| | BOC | 974.28 | 888.44 | 828.09 | 748.63 | 712.56 | 615.10 | 509.38 | **753.79** | 888.99 | 654.81 | 664.69 | **736.16** |
| | RBOC | 990.88 | 922.69 | 827.79 | 743.90 | 685.01 | 570.73 | 452.77 | **741.97** | 871.02 | 679.75 | 720.73 | **757.17** |
| CO2 Emission | CC | 3081.75 | 3224.37 | 3203.08 | 3138.05 | 3164.54 | 3160.71 | 3094.34 | **3152.41** | 3197.32 | 3095.59 | 3124.18 | **3139.03** |
| | BOC | 2330.21 | 1987.46 | 2154.55 | 1420.60 | 1971.07 | 2001.15 | 1517.49 | **1911.79** | 2408.85 | 2135.09 | 2382.58 | **2308.84** |
| | RBOC | 2437.29 | 2535.05 | 2175.94 | 1644.04 | 2026.19 | 2395.38 | 1599.73 | **2116.23** | 2390.89 | 1931.84 | 2299.76 | **2207.50** |
| Cooling | CC | 1884.81 | 1662.41 | 1431.68 | 1282.45 | 1076.87 | 980.00 | 793.48 | **1301.67** | 1536.47 | 1044.85 | 1051.21 | **1210.84** |
| | BOC | 1909.97 | 1683.69 | 1567.54 | 1346.26 | 1249.09 | 1059.42 | 852.95 | **1381.28** | 1702.81 | 1196.38 | 1159.99 | **1353.06** |
| | RBOC | 2000.37 | 1774.87 | 1549.64 | 1326.55 | 1219.00 | 1001.70 | 794.71 | **1380.98** | 1670.63 | 1226.62 | 1291.51 | **1396.25** |